\documentclass{aa}  
\usepackage{graphicx}
\usepackage{txfonts}
\usepackage{natbib}
\usepackage{color}
\usepackage{hyperref}
\usepackage{nccmath}
\hypersetup{colorlinks=true,allcolors=[rgb]{0,0,0.8}}

\usepackage{showyourwork}

\makeatletter
\renewcommand*\aa@pageof{, page \thepage{} of \pageref*{LastPage}}

\def\kms{\ifmmode{\rm km\th s^{-1}}\else km\th s$^{-1}$\fi}
\def\th{\thinspace}

\makeatother
\begin{document} 

\title{The eclipse of the V773 Tau B circumbinary disk\thanks{The manuscript and processing scripts are available in a GitHub repository at \url{https://github.com/mkenworthy/V773TauBdisk}. Processed photometric data and light curves are available in electronic form at the CDS via anonymous ftp to \url{cdsarc.u-strasbg.fr} (130.79.128.5) or via \url{http://cdsweb.u-strasbg.fr/cgi-bin/qcat?J/A+A/}}}

   \author{M.A. Kenworthy\inst{1}
          \and
          D. Gonz\'{a}lez Picos\inst{1}
          \and
          E. Elizondo\inst{2}
          \and
          R.G. Martin\inst{3}
          \and
          D.M. van Dam\inst{1}
          \and
          J.E. Rodriguez\inst{4}
          \and
          G.M. Kennedy\inst{5}
          \and
          C. Ginski\inst{1}
          \and
          M. Mugrauer\inst{6}
          \and
          N. Vogt\inst{7}
          \and 
          C. Adam\inst{8}
          \and
         R.J. Oelkers\inst{9}
}
   \institute{Leiden Observatory, University of Leiden,
   PO Box 9513, 2300 RA Leiden, The Netherlands\\
   \email{kenworthy@strw.leidenuniv.nl}
    \and
    Department of Physics and Astronomy, Wayne State University, 666 W Hancock St, Detroit, MI 48201, USA
    \and
    Department of Physics and Astronomy, University of Nevada, Las Vegas, 4505 South Maryland Parkway, NV 89154, USA
    \and 
    Department of Physics and Astronomy, Michigan State University, East Lansing, MI 48824, USA
    \and
    Department of Physics, University of Warwick, Coventry CV4 7AL, UK
    \and
    Astrophysikalisches Institut und Universitäts-Sternwarte Jena, Schillergäßchen 2, D-07745 Jena, Germany
    \and
    Instituto de F\'{i}sica y Astronom\'{i}a, Facultad de Ciencias, Universidad de Valpara\'{i}so, Av. Gran Breta\~{n}a 1111, Playa Ancha, Valpara\'{i}so, Chile
\and
    Centro de Astronom\'ia (CITEVA), Universidad de Antofagasta, Avenida U. de Antofagasta, 02800, Antofagasta, Chile
    \and
    Munnerlyn Astronomical Instrumentation Laboratory, Department of Physics and Astronomy, Texas A\&M university, College Station, TX 77843 USA
    }

   \date{Received 2022-02-28; accepted 2022-06-14}

 
  \abstract
   {Young multiple stellar systems can host both circumstellar and circumbinary disks composed of gas and dust, and the orientations of circumbinary disks can be sculpted by the orientation and eccentricity of the central binaries.
   Studying multiple binary systems and their associated disks enables understanding of the size and distribution of the planetary systems that subsequently form around them.}
   {A deep ($\sim 70$\%) and extended ($\sim$150 days) eclipse was seen towards the young multiple stellar system V773 Tau in 2010.
   We interpret it as due to the passage of a circumbinary disk around the B components moving in front of the A components.
 Our aim is to characterise the orientation and structure of the disk, to refine the orbits of the subcomponents, and to predict when the next eclipse will occur.}
   {We combine the photometry from several ground based surveys, construct a model for the light curve of the eclipse, and use high angular resolution imaging to refine the orbits of the three components of the system, A, B and C.
   Frequency analysis of the light curves, including from the TESS satellite, enables characterisation of the rotational periods of the Aa and Ab stars.}
   {A toy model of the circumbinary disk shows that it extends out to approximately 5 au around the B binary and has an inclination of 73\degr{} with respect to the orbital plane of AB, where the lower bound of the radius of the disk is constrained by the geometry of the AB orbit and the upper bound is set by the stability of the disk.
   We identify several frequencies in the photometric data that we attribute to rotational modulation of the Aa and Ab stellar companions.
   We produce the first determination of the orbit of the more distant C component around the AB system and limit its inclination to 93\degr{}.
}
   {The high inclination and large diameter of the disk, together with the match from theory suggest that B is an almost equal mass, moderately eccentric binary.
   We identify the rotational periods of the Aa and Ab stars, identify a third frequency in the light curve that we attribute to the orbital period of the stars in the B binary.
   We predict that the next eclipse will be around 2037, during which both detailed photometric and spectroscopic monitoring will characterise the disk in greater detail.}


   \maketitle
%

\section{Introduction}

Young multiple stellar systems are common \citep[e.g.][]{Ghez93,Duchene13} as are gas disks in and around the stellar components \citep{Akeson19}.
Chaotic accretion can occur during the star formation process as a result of turbulence within the molecular cloud from which the stars form \citep{Bate2003,McKee2007}.
Circumstellar and circumbinary disks can form with a misalignment to the binary orbital plane \citep{Monin2007, Bate2018}.

These binaries have been shown to influence and sculpt the circumbinary disks around them, possibly reinvigorating planet formation \citep{Cabrit06,Rodriguez18}.
The torque from a binary or multiple stellar system affects the formation and evolution of gas disks \citep[e.g.][]{Nelson2000, Mayer2005,Boss2006, Fu2017} and the interaction of planets with the disk \citep{Picogna2015,Lubow2016,Martin2016}.
Thus understanding how stellar multiplicity affects the formation and evolution of disks is important to explain observed exoplanet properties.

A misaligned circumbinary disk undergoes retrograde nodal precession.
If the binary is in a circular orbit, the precession is always around the binary angular momentum vector \citep[e.g.][]{Larwood1996}.
However, around an eccentric binary the precession is around the binary eccentricity vector if the initial misalignment is sufficiently high \citep{Farago2010,Aly2015}.
Dissipation within the disk causes the disk to align towards coplanar to the binary orbit, or polar to the binary orbit and aligned to the binary eccentricity vector \citep{Martin17,Lubow2018,Zanazzi2018}.
While all currently detected circumbinary planets are in coplanar orbits, around an eccentric binary polar planets may be more stable than coplanar planets \citep{Chen2020} and terrestrial planets at least form more efficiently \citep{Childs2021}.
If the disk lifetime is shorter than the disk alignment timescale then planets may form in misaligned orbits.

There are now many observations of circumbinary disks with a range of misalignment angles.
For example, KH~15D has a misalignment of about $3-15^\circ$ \citep{Chiang2004, Winn2004,Capelo2012,Poon2021}.
IRS~43 is misaligned by greater than $60^\circ$ \citep{Brinch16}, and the circumbinary disk around 99 Her has a polar misalignment \citep{Kennedy12}.
HD~98800 has a misalignment of $90^\circ$, meaning it is polar aligned to the central binary orbit \citep{Kennedy19,Zuniga-Fernandez21}.
Of the two possible disk inclinations allowed by observations, the polar alignment was first suggested by considering the dynamics of a disk since the polar alignment timescale is short compared to the stellar age.
Furthermore, the size of the inner hole in the disk that is carved by the binary orbit \citep{Artymowicz1994} is in agreement with a polar alignment \citep{Franchini19}.
There is also an external binary companion to the circumbinary disk in this system and its orbit is close to $34^\circ$ to the disk angular momentum.

V773 Tau is a young multiple stellar system that has been intensively studied as described in Section~\ref{sec:v773}.
It contains three components, each at a distinctly different stage of young stellar evolution - the A component is an almost equal mass binary, which in itself orbits around an almost equal mass B component that has an edge-on disk around it, forming a hierarchical multiple system with a decades long orbit.
Both these components are orbited by a third, C component that is highly embedded within a cloud of dust and is on a several hundred year orbit around both A and B.

In Section~\ref{sec:data} we present the photometry of the V773 system from several ground based surveys.
A deep extended eclipse occurs over a 200 day period, and we attribute this to a disk around the B component passing in front of the A components.
This disk is highly inclined with respect to the AB orbital plane, which should otherwise become coplanar within a dynamically short timescale.
Previous work \citep{Boden12} indicates that the B component is itself a tight, almost equal mass binary system.
If the B binary is also eccentric, then the disk will experience a torque that keeps it at the inclination that we observe.
Time series analysis of the photometry reveals two rotational periods consistent with the stars in the A binary, and a third period of approximately 67 days is revealed in the analysis.
If this period is attributed to the orbital period of the B binary, then we show that a moderate orbital eccentricity will keep the circumbinary disk at the observed inclination.
We derive the properties of the B binary and the circumbinary disk, and in Section~\ref{sec:orbitalfit} use the astrometry from the direct imaging observations to determine the expected epoch of the next eclipse of the A system.

Section~\ref{sec:variability} details our analysis of the photometric fluctuations due to both rotational modulation of Aa and Ab, and identifies a photometric signal whose period is a plausible measure of the orbit of the BaBb binary.
The long and extended eclipse is modeled in Section~\ref{sec:model} as an azimuthally symmetric dusty disk that is highly inclined to the AB orbital plane.
Theoretical consideration of the torques from an eccentric binary on a surrounding disk are discussed in Section~\ref{sec:theory} along with the prediction for the timing of the next eclipse in Section~\ref{sec:discuss}.
We summarise our results and discuss future observations in Section~\ref{sec:conclusions}.

\section{The V773 Tau system}\label{sec:v773}

\begin{figure}
\begin{center}
    \centering
    \includegraphics[width=\columnwidth]{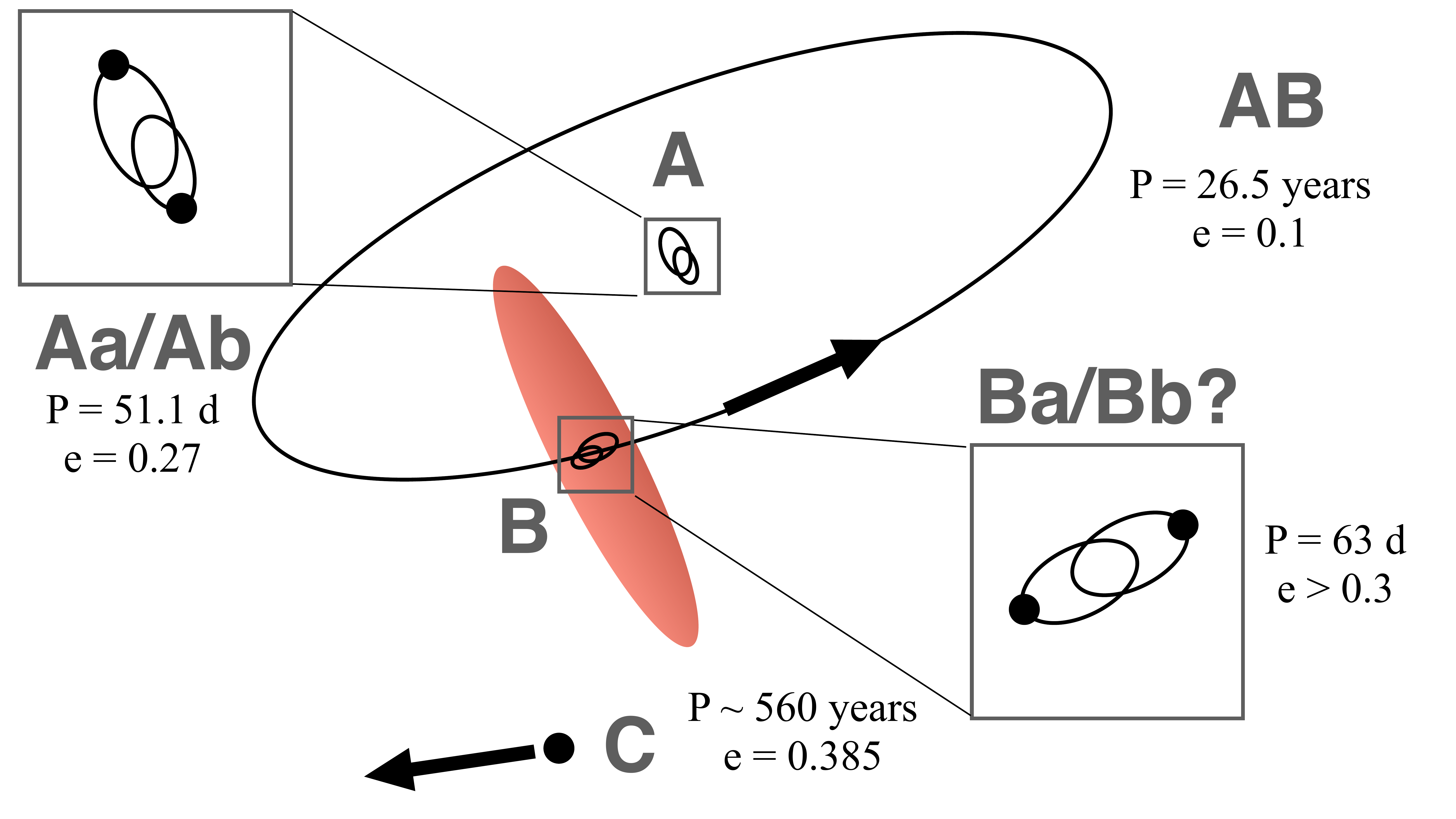}
    \caption{Sketch of the V773 Tau system. 
    We hypothesise that B is itself an equal mass binary system with a moderately eccentric orbit.
    Orbits are not to scale.}
\label{fig:cartoon}
\end{center}
\end{figure}

\begin{figure*}
\begin{center}
    \centering
    \includegraphics[width=\textwidth]{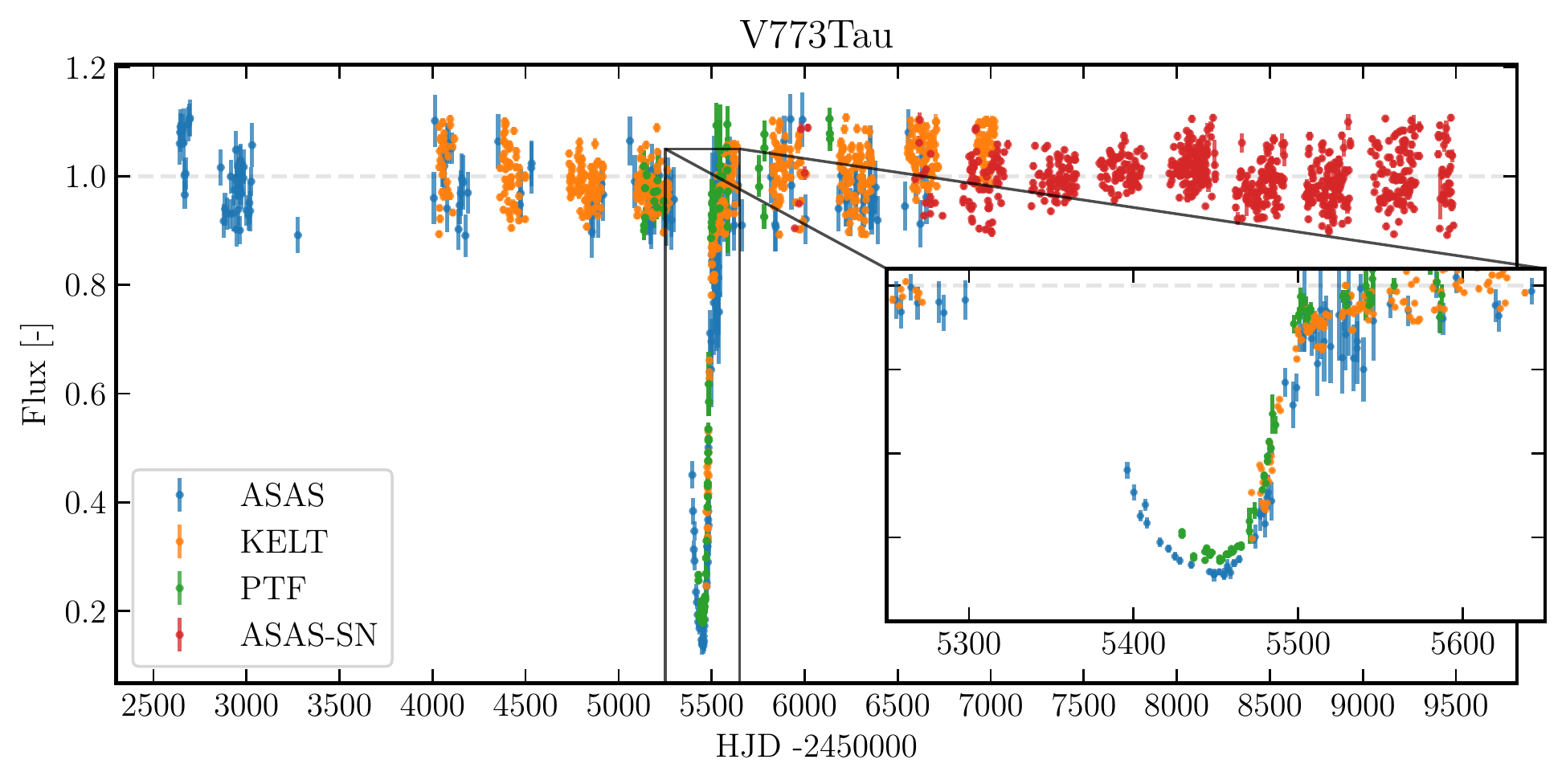}
    \caption{Optical photometry of the V773 Tau system.
    The four different photometric surveys are represented by different colours, and each photometric series has been normalised using flux outside the main eclipse.
    The inset panel shows the eclipse light curve in greater detail.
    The PTF and KELT curves show different depths of the eclipse, indicating the presence of sub-micron dust that causes different amounts of optical extinction in the different pass bands of the two surveys.}
  \label{fig:lc}
  \script{fig2_complete_lightcurve.py}
\end{center}
\end{figure*}

\subsection{Discovery and architecture}

V773 Tau (HD 283447, HBC 367) is a young \citep[$3\pm1$ Myr;][]{Boden07} stellar system that was identified as a young T Tauri star \citep{Rydgren76} and subsequently resolved as a visual double \citep{Ghez93,Leinert93}.
Its age was determined by comparing the temperature and luminosity of the stars Aa and Ab with PMS stellar models from \citet{Montalban06}, which show excellent consistency with the component parameters from \citet{Boden07}.
It is located at a distance of $132.8\pm 2.3 $ pc as determined by orbital and trigonometric parallax \citep{Torres12}, in contradiction with the larger distance determined by Gaia EDR3.
We attribute this discrepancy to the multiple source nature of the system which is not resolved by GAIA, and we use the distance determined by the spatially resolved A components with the VLA observations \citep{Torres12}.

The bound components of V773 Tau were initially identified as A, B and C (see Figure~\ref{fig:cartoon}).
The A component was resolved by Very Long Baseline Array observations to be a binary system with $P_{AaAb}=51.1003\pm 0.0022$ days and an eccentricity of $e=0.2710\pm 0.0072$ \citep{Torres12}.
The A binary becomes more luminous in radio waves during the periastron passage of the A stars, which is thought to be from the interacting magnetic fields of the two stars \citep{Torres12}.

A is orbited by the B component on a period of approximately 26 years, and both A and B are orbited by a C component with an orbital period of several hundred years \citep{Duchene03}.
C is a heavily reddened companion that is the faintest component at a wavelength of two microns but is the brightest at $4.7\mu m$ \citep{Duchene03,Woitas03}.
The flux contribution from C at optical wavelengths is approximately 100 times fainter than the A and B components \citep{Duchene03}.

\subsection{The B component}

Photometry of the B component shows a significantly greater amount of extinction compared to the A component, implying the presence of additional dust along the line of sight to the B component \citep{Torres12}.
The B component shows significant photometric variability on the order of 2 to 3 magnitudes in the $K$ band \citep{Boden12}, consistent with clouds or clumps of dust orbiting around the B component.
The initial fit to the spectral energy distribution (SED) suggested that the B component consisted of a single K star \citep{Duchene03}, but subsequent radial velocity monitoring and astrometry of the A binary enabled a dynamical mass determination of both the B and A components, revealing a much larger mass of $2.35 M_\odot$ for the B component \citep{Boden12}.
A single $2.35 M_\odot$ star would have a luminosity of $17 L_\odot$, inconsistent with the optical luminosity observed and the extinction derived.
If the B system is itself a binary, BaBb, then assuming equal mass components of $1.5 M_\odot$ yields a much lower luminosity that is consistent with the observed SED.
We therefore assume that the V773 Tau AB system is a hierarchical quadruple system, with a circumbinary disk around the B component which responsible for the significant reddening and photometric variability seen towards the B component.

%

\section{Data}\label{sec:data}

\subsection{Discovery of the Eclipse with KELT}

The Kilodegree Extremely Little Telescope \citep[KELT; ][]{Pepper07,Pepper12, Pepper2018} survey is a project searching for transiting exoplanets, using two robotic wide-field telescopes, one at the Winer Observatory in Arizona (KELT-N) and the Southern African Astronomical Observatory (SAAO) near Sutherland (KELT-S).
The observatories consist of a $4096\times 4096$ AP16E Apogee CCD camera (KELT-N) and $4096\times 4096$ Alta U16M Apogee CCD camera (KELT-S) using a Kodak Wratten \#8 red-mass filter, comparable to an $R$ band magnitude.
The cameras are fed with a $f/19$ 42 mm (wide angle survey mode with 26\degr{}$\times$26\degr{} field of view) or a $f/28$ 200 mm (narrow angle campaign mode with 10.8\degr{}$\times$ 10.8\degr{} field of view) Mamiya lenses.
Images are obtained at a cadence of 10 to 30 minutes and yield useful magnitudes from 7 to 13.

\begin{table*}
\caption{Orbital parameters for A and B systems. Data from \citet{Torres12} ``Joint solution'' for AaAb and from \citet{Boden12} for A-B. NOTE $\omega_A$ is defined in the RV sense for Torres and Boden, so that you will need to add 180\degr{} to $\omega_A$ for the Torres and Boden orbits to be correctly calculated in orbitize!}              
\label{tab:orbs}      
\centering                                      
\begin{tabular}{l c c c c}          
\hline\hline                        
Orbital  & Torres 2012  & Boden 2012    & this work   &  this work \\
 Parameter & AaAb  & AB    & AB   & AB-C  \\

\hline                                   
Period          & $51.1003\pm 0.0022$ d  & $26.20 \pm 1.1$  yr  &  $26.50  \pm 0.07$  yr  & $624_{-52}^{+83}$ yr\\
$T_0$           & $53059.92\pm 0.33$ (MJD) & $2010.53 \pm 1.0$ yr &  $2010.11\pm 0.12$  yr  & $1740_{-40}^{+26}$\\
$e$             & $0.2710\pm 0.0072$    & $0.099 \pm  0.026$ &  $0.104  \pm 0.009$    & $0.40_{-0.05}^{+0.04}$ \\
$\omega_A$ (deg)& $5.2\pm 2.7$          & $94 \pm 17$        &  $266.0 \pm 1.5$    & $107.8\pm 4.5$ \\
$\Omega$ (deg)  & $63.3\pm 1.1$         & $288.2 \pm 1.0$    &  $290.54 \pm 0.40$    & $104.6\pm 1.3$\\
$i$ (deg)       & $67.6\pm 1.5$         & $71.48 \pm 0.78$   &  $69.25 \pm 0.40 $    & $97.3\pm 0.6$\\
$a$ (mas)       & $2.837\pm 0.035$      & $115.5 \pm 3.4$    &  $117.7 \pm 0.86$    & $1013_{-74}^{+93}$\\
\hline                                             
\end{tabular}
\end{table*}

\begin{table}
\caption{Physical parameters for V773 Tau A-B. Data from \citet{Torres12} ``Joint solution'' for AaAb and from \citet{Boden12} for A-B}              
\label{tab:physparams}      
\centering                                      
\begin{tabular}{l c c}          
\hline\hline                        
Parameter & Value & Note \\
\hline                  
System distance (pc)        & $132.8 \pm 2.4$     & T2012 \\
A-subsystem mass ($M_\odot$)& $2.91 \pm 0.20$     & B2007; T2012 \\
A–B semimajor axis (AU)     & $15.35 \pm 0.45$    & B2012 \\
A–B system mass ($M_\odot$) & $5.27 \pm 0.65$     & B2012 \\
B mass ($M_\odot$)          & $2.35 \pm 0.67$     & B2012 \\
\hline                                             
\end{tabular}
\end{table}

The KELT survey has been highly successful in discovery and analysis of systems being eclipsed by circumstellar material, including the longest period eclipsing binary \citep{Rodriguez:2016}.
To search for eclipses of YSOs, we cross-matched the \citet{Zari2018} catalog of young stars in the solar neighbourhood that were identified by a combination of Gaia DR2 \citep{GaiaDR2} photometry and kinematics to the KELT catalog to obtain a sample of $\sim$1450 stars with lightcurves.
The corresponding KELT lightcurves for each YSO target were visually inspected for any large dimming events ($>$10\%) that lasted more than a week.
From this analysis, we identified a $\sim$80\% deep dimming event in V773 Tau, shown in Figure~\ref{fig:lc}.
The eclipse is incomplete, with the first half of the eclipse occurring when V773 Tau was behind the Sun, so we only observed the egress of the eclipse that lasted $\sim$100 days.
Based on this discovery, we searched for the eclipse in other photometric data sets.

\subsection{ASAS}

The All Sky Automated Survey \citep[ASAS; ][]{pojmanski_all_1997, asas_2005, asas_2018} is a survey consisting of two observing stations - one in Las
Campanas, Chile and the other on Maui, Hawaii. 
Each observatory is equipped with two CCD cameras using V and I filters and commercial f $ = 200$ mm, D $= 100$ mm lenses, although both larger (D $= 250$ mm) and smaller (50-72 mm) lenses were used at earlier times.
The majority of the data are taken with a pixel scale of $\approx$ 15\arcsec{}.
ASAS splits the sky into 709 partially overlapping (9\degr{} $\times$ 9\degr{} fields, taking on average 150 3-minute exposures per night, leading to a variable cadence of 0.3-2 frames per night.
Depending on the equipment used and the mode of operation, the ASAS limiting magnitude varied between 13.5 and 15.5 mag in V, and the saturation limit was 5.5 to 7.5 mag. 
Precision is around 0.01-0.02 mag for bright stars and below 0.3 mag for the fainter ones. 
ASAS photometry is calibrated against the Tycho catalog, and its accuracy is limited to 0.05 mag for bright, non-blended stars.

 \begin{table*}
\caption{Summary of the photometric data for V773 Tau employed in this work. Surveys with a * observed the eclipse. The start and end date are in HJD-2450000 format.}              
\label{tab:v773tau_photometry}      
\centering                                      
\begin{tabular}{l c c c}          
\hline\hline                        
Survey  & Start Date & End Date & N$_{\text{points}}$ \\
\hline                  
ASAS*              & 2621.65 &  8879.56& 712\\
SWASP  & 3215.71 & 4542.35 &  4902\\
HATNET    &4390.97 & 4552.81 & 3402\\
PTF*     & 5543.16 & 5782.49 & 121\\
ASAS-SN   & 5946.95 & 8451.85 & 882\\
KELT*   & 7020.84 & 2986.11 & 10272\\
ASAS-SN   & 8002.90 & 9488.90  & 1630\\
TESS      & 8816.10 & 8840.87 & 999\\
\hline
\end{tabular}
\end{table*}

\subsection{ASAS-SN}

The All Sky Automated Survey for Supernovae \citep[ASAS-SN; ][]{shappee_man_2014,kochanek_all-sky_2017} consists of six stations around the globe, with each station hosting four telescopes with a shared mount.
The telescopes consist of a 14-cm aperture telephoto lens with a field of view of approximately 4.5\degr{}$\times$4.5\degr{} and an 8.0\arcsec{} pixel scale.
Two of the original stations (one in Hawaii and one in Chile) are fitted with $V$ band filters, whereas the other stations (Chile, Texas, South Africa and China) are fitted with $g$ band filters.
ASAS-SN observes the whole sky every night with a limiting magnitude of about 17 mag in the $V$ and $g$ bands.

\subsection{PTF}

The Palomar Transient Factory (PTF) is an automated wide field optical photometric survey described in \citet{Law09}.
It uses an 8.1 square degree camera with 101 megapixels at  1\arcsec\ sampling mounted on the 48 inch Samuel Oschin telescope at the Palomar Observatory.
Nearly all of the images are taken in one of two filters - Mould-$R$ and SDSS-$g'$, reaching $m_{g'}\approx 21.3$ and $m_R \approx 20.6$ in 60 s exposures.
Over 120 photometric observations were obtained using the $R$ band filter \citep{Ofek12}.

\subsection{TESS}

The Transiting Exoplanet Survey Satellite \citep[TESS; ][]{2015JATIS...1a4003R} is a satellite designed to survey for transiting exoplanets among the brightest and nearest stars over most of the sky.
The TESS satellite orbits the Earth every 13.7 days on a highly elliptical orbit, scanning a sector of the sky spanning 24\degr $\times$ 96\degr\ for a total of two orbits, before moving on to the next sector. 
It captures images at a 2 second (used for guiding), 20 seconds (for 1 000 bright asteroseismology targets), 120 seconds (for 200 000 stars that are likely planet hosts) and 30 minutes (full frame image) cadences.
The instrument consists of 4 CCDs each with a field of view of 24\degr$\times$24\degr, with a wide band-pass filter from 600-1000 nm (similar to the $I_C$ band) and has a limiting magnitude of about 14-15 mag ($I_C$).
The data was extracted from the TESS archive using the {\tt eleanor} package \citep{Feinstein19} which corrected for known systematics in the cameras and telescope.

\subsection{Combined light curve}

A summary of all the photometry obtained for V773 Tau is shown in Table~\ref{tab:v773tau_photometry}.
Photometric data from each survey is normalised to account for camera and instrumental throughput offsets.
The relative flux is computed for each subset of points, where the baseline is determined by ignoring in-eclipse data.
The data out of eclipse is resampled to 1-day bins to improve the signal-to-noise ratio and reduce the number of points for display and analysis.
Photometric outliers are removed by sigma-clipping on the flux and rejecting points with large photometric error values.
Further processing includes removing the rotational variability due to the presence of spots on the A components, which is removed for each survey by fitting a stellar variability model (see Section~\ref{sec:variability}). 

The light curve of V773 Tau is shown in Figure~\ref{fig:lc} with the photometry from all four ground-based surveys.
The depth of the eclipse is approximately 80\% at MJD~55450, with an estimated full width at half minimum of approximately 100 days. 
The start of the eclipse was not observed due to the observing season for the star.
Three surveys show the eclipse, with the PTF survey showing a slightly shallower eclipse depth compared to the ASAS light curve.
The PTF observed V773 Tau in the $R$-band while ASAS observations were in the $V$-band, as a result there is a color depth difference due to the wavelength dependent nature of dust absorption, indicating sub-micron sized dust.
The data from TESS is shown in Figure~\ref{fig:tess_variability}.

\subsection{Adaptive optics imaging}\label{sec:di}

\begin{figure}[ht]
    \centering
    \includegraphics[width=\columnwidth]{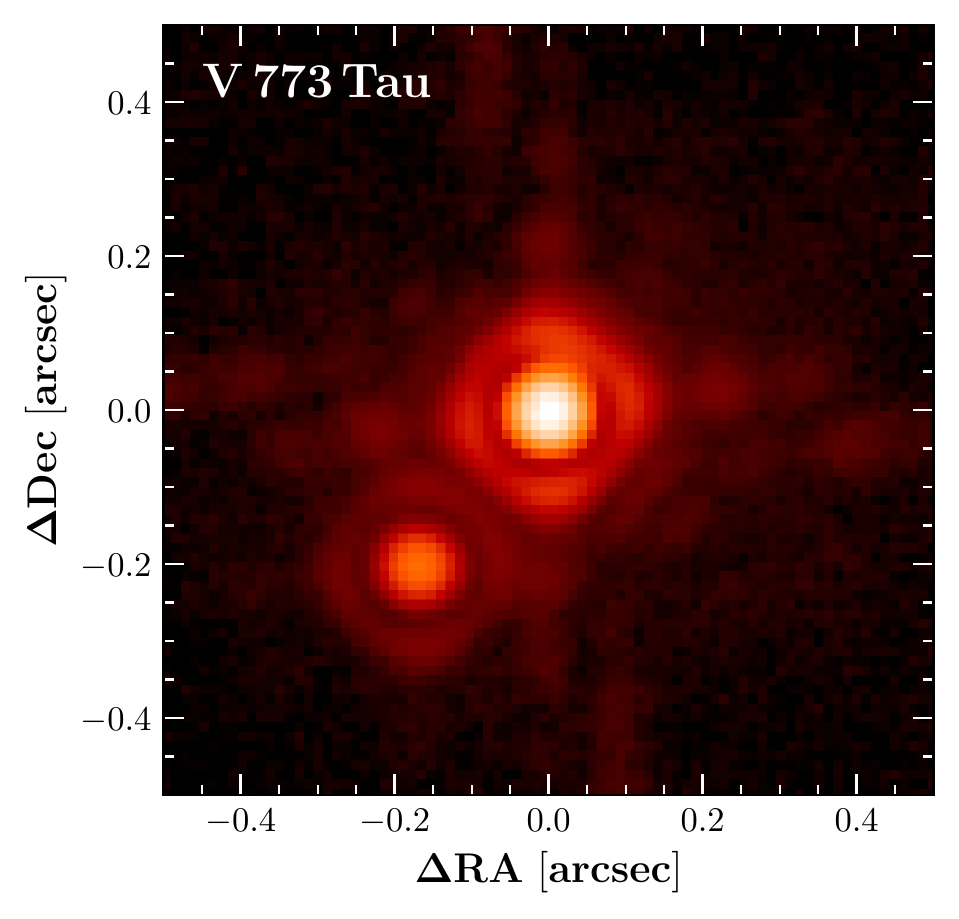}
    \caption{SPHERE image of V773 Tau AB and C. The axes are centred on A, and B is not visible in this image, being under the first diffraction ring of the A component.
    C is shown to the lower left of AB.
    North is up and East to the left.}
  \label{fig:sphere}
  \script{make_fig3_v773tau.py}
\end{figure}

The V773 Tau system was observed with the SPHERE/IRDIS \citep[Infra-Red Dual Imaging and Spectrograph; ][]{Beuzit19,Dohlen08}, mounted at the Nasmyth platform of the Unit 3 telescope (UT3) at ESO's VLT on 2021 Oct 28 (MJD 59515).
The IRDIS camera was used to obtain 16 images with 16s integration in direct imaging mode with no coronagraphs for a total integration time of 256 seconds.
The astrometric PSF extraction was done by simultaneously fitting 2 elliptical Moffat functions to the flux image.
The official astrometric calibration for the $BB-K$ filter \citep{Maire16} was used to correct the observations.
The image was corrected for geometric distortions and the true north offset (see Figure~\ref{fig:sphere}).
For the parallactic angle the header values of the data were used. 
Four independent measurements of the PA and separation were made (left and right detector side, beginning and end of the sequence). 
The resultant position is listed in Table~\ref{tab:newastrom}.

\begin{table*}
\caption{Orbital measurements for A, B and C.}              
\label{tab:newastrom}      
\centering                                      
\begin{tabular}{l c c c c c c c c}          
\hline\hline
 &     $\theta(AB)$ & $\sigma_{\theta(AB)}$ & $\rho(AB)$ & $\sigma_{\rho(AB)}$ & $\theta(AC)$ & $\sigma_{\theta(AC)}$ & $\rho(AC)$ & $\sigma_{\rho(AC)}$ \\
    &     ($\degr$) & ($\degr$)             & (mas)      & (mas)               & ($\degr$)    & ($\degr$)             & (mas)      & (mas)               \\
\hline
Moffat fit &     -- &                --     &   --       &  --                 & 139.94          & 1.1                   & 264.6      & 6.8                 \\
Orbitize!  &  339.1 & 0.5                   & 73.8       & 0.6                 & 139.92          & 0.04                  & 262.05     & 0.28                \\
\hline
\end{tabular}
\end{table*}

To confirm the astrometry of AC and to measure the AB separation in the presence of A's diffraction rings, a custom fitting routine was used.
The centre of A and C was determined by choosing a trial value for the $(x,y)$ position of the light centroid of the stellar PSF, then rotating by 180\degr{} and subtracting this image from the unrotated image.
The  {\tt emcee} package \citep{foreman-mackey2013} was then used to minimise the calculated $\chi^2$ of the residuals within a disk with diameter equal to the Airy disk of the PSF centered on the trial coordinates, and the pixel positions along with the errors derived through marginalisation of the distributions was calculated.

\begin{figure}
\begin{center}
    \centering
    \includegraphics[width=\columnwidth]{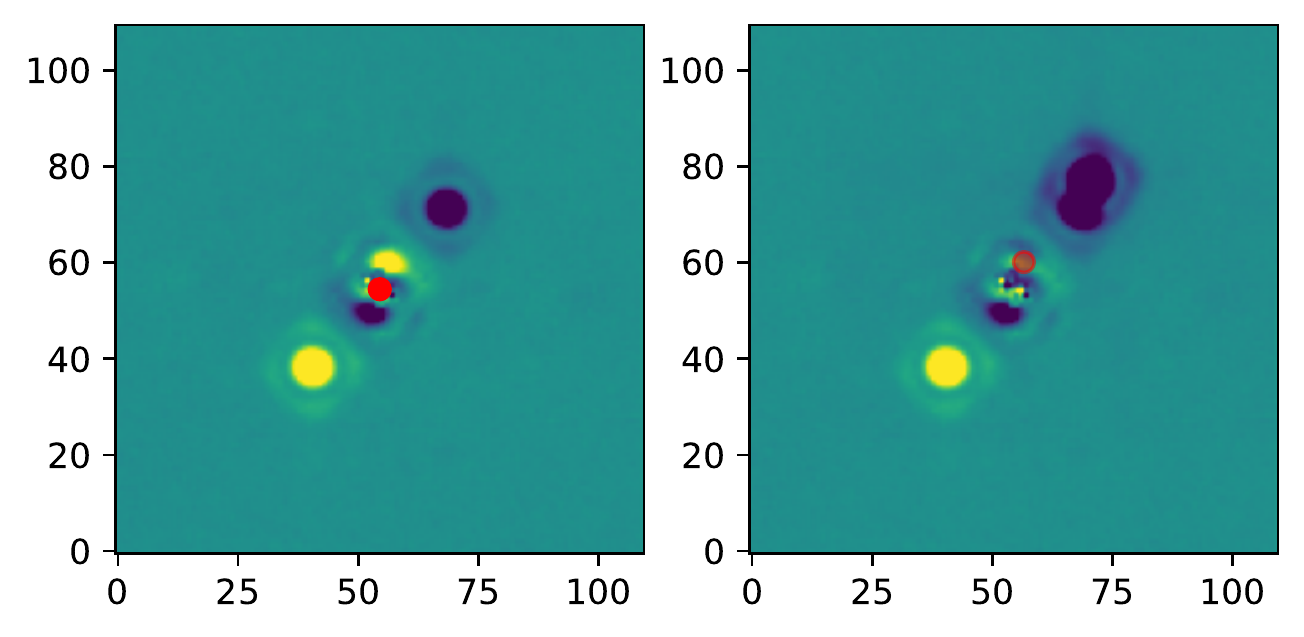}
    \caption{Determining the astrometry of V773 Tau A, B and C.
    The left hand panel shows the rotation subtracted image of the V773 system, which was constructed by rotating the image by 180 degrees around the centroid of A (indicated by the red dot), then subtracting it from the original image.
    The positive flux image of C is seen to the lower left and the negative flux image of C to the upper right.
    The positive flux image of the B component is then seen to the upper right of the centroid of A, indicating its location.
    The right hand panel shows the subtraction of the positive image of C from the location of B, with B marked with a red circle.
    Residuals from the subtraction processes can be seen around the location of A and B.}
  \label{fig:rotsub}
  \script{make_astrometry_ABC_plot.py}
\end{center}
\end{figure}

With the centroid of star A determined, we then subtracted a 180 degree rotated image of A from the original image, revealing the B system at the location of the first diffraction ring of system A (see Figure~\ref{fig:rotsub} left panel).
We then use star C as a reference PSF, translate it to the location of B, scale the flux of the image by a factor $f$ and then use {\tt emcee} to determine the relative separation of B and C (see Figure~\ref{fig:rotsub} right panel).
With the absolute pixel positions of A and C, the relative position of B with respect to C and the pixel scale of the SPHERE IRDIS camera, the position angles and separations of AB and AC are determined. 
The AC values are found to be consistent with the Moffat fitting procedure for AC and all fitting results are listed in Table~\ref{tab:newastrom}.

Errors reported on the astrometry using {\tt emcee} are smaller than those reported in previous measurements on larger telescopes, even though our signal to noise is similar.
The fitting routine does not include systematic errors, which are difficult to quantify with the diffraction ring residuals seen at the location of the B component.
We therefore doubled the errors on the measured relative position of A from B, which is consistent with the errors reported by other papers, to represent the systematic errors in our fitting.

\section{Orbital fitting for A, B and C}\label{sec:orbitalfit}

\begin{figure*}[ht]
\begin{center}
    \centering
    \includegraphics[width=\textwidth]{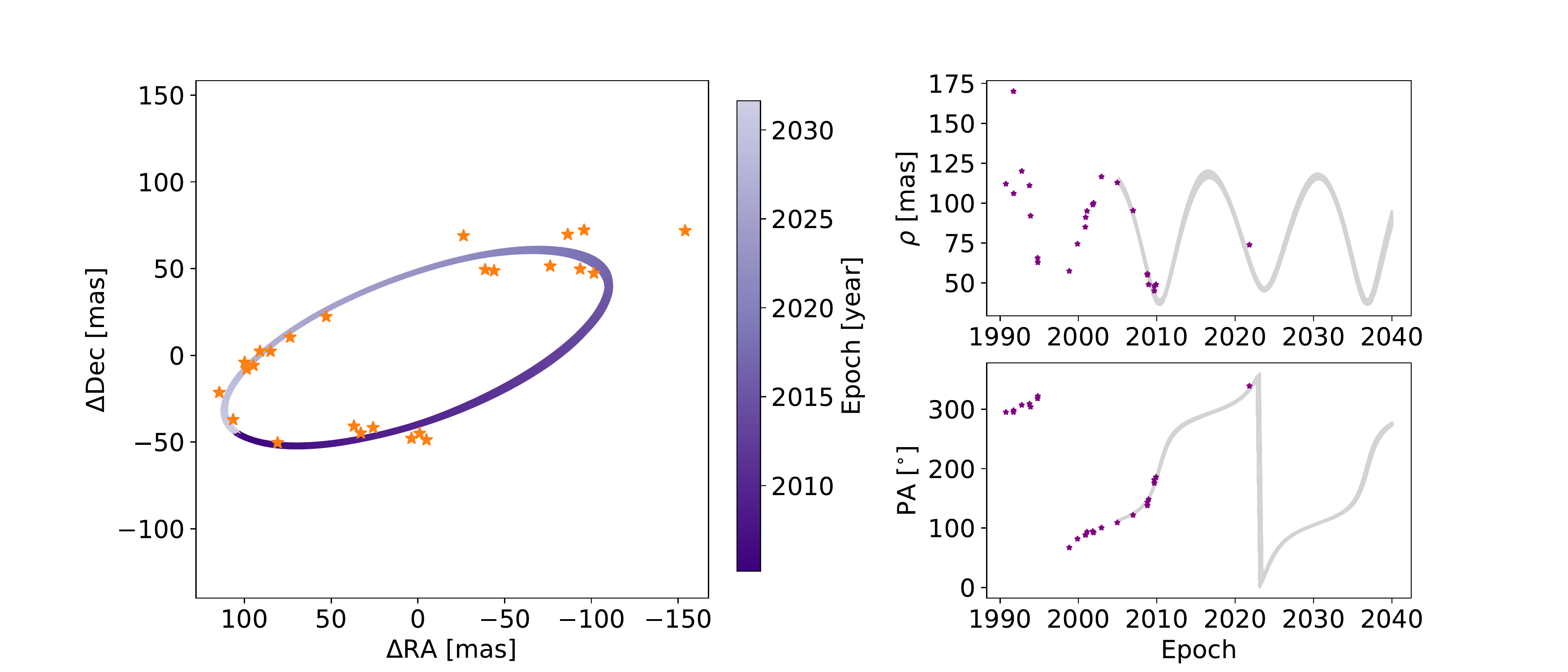}
        \caption{The orbit of the V773 Tau AB system.
        A is fixed at the origin in the left hand panel.
        The right hand panels show the separation and position angles of the astrometry.}
      \label{fig:v773orbitize}
      \script{analyze_and_plot_AB_orbit_bundle.py}
\end{center}
\end{figure*}

\begin{figure*}[ht]
\begin{center}
    \centering
    \includegraphics[width=\textwidth]{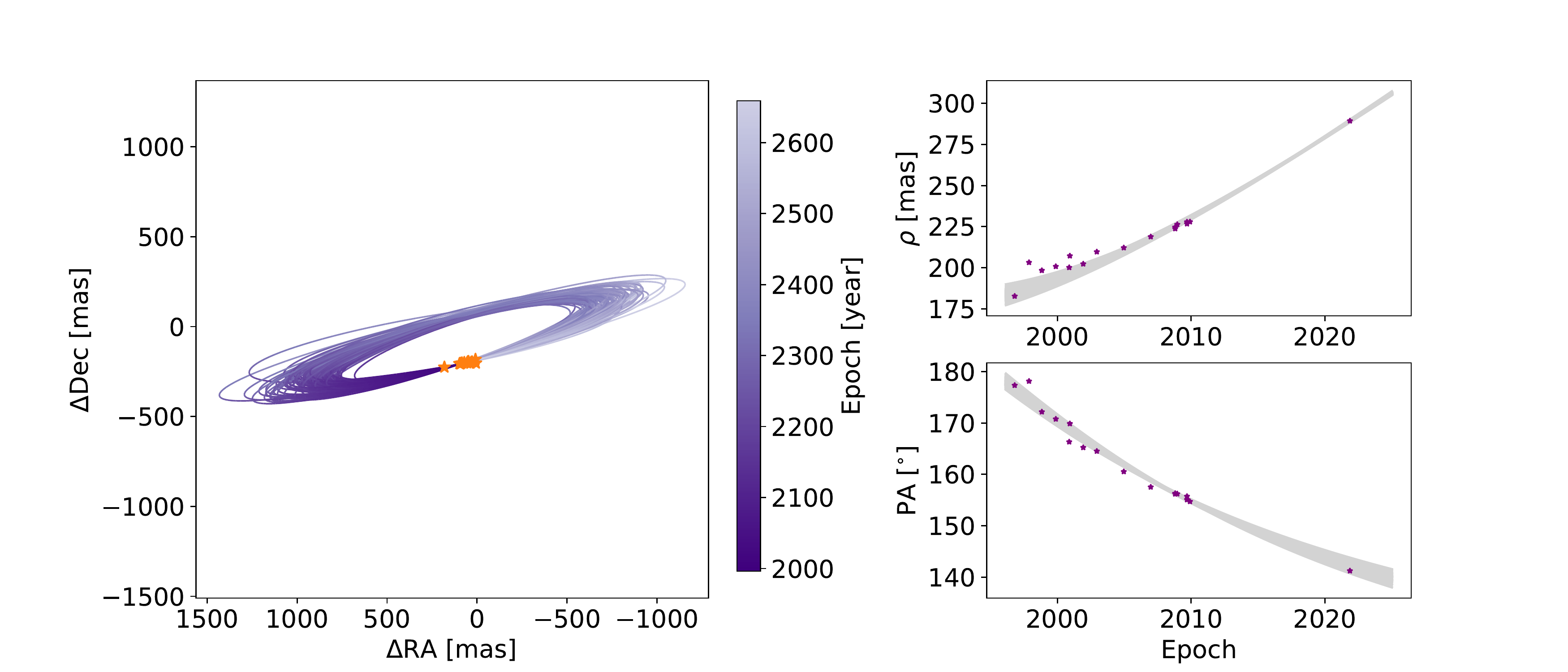}
        \caption{The orbit of the V773 Tau C system.
        The barycentre of AB is fixed at the origin in the left hand panel.
        The right hand panels show the separation and position angles of the astrometry.}
    \label{fig:v773Corbitize}
    \script{analyze_and_plot_C_orbit_bundle.py}
\end{center}
\end{figure*}

We combined previously obtained astrometry of the V773 Tau system \citep[from the analysis of ][]{Duchene03,Boden12} with our reported astrometry, and used the {\tt orbitize!} \citep{Blunt20,foreman-mackey2013} package to perform an updated orbital fit for the AB system and the orbit of C around the barycentre of AB.
For {\tt orbitize!} we use 1000 walkers, the number of temperatures is 20 (to move walkers out of local minima in the optimization function), a burn in of 2000 per walker and a total run of $10^6$ steps for both systems.
The output is in the form of orbits (consisting of six orbital elements, the total mass of the system and the parallax) with fits consistent with the astrometry, which are referred to as a `bundle'.
By marginalising over all the other orbital parameters in the bundle, a distribution for each orbital parameter is constructed.
The value and quoted errors for the orbital elements are at the 16th, 50th and 84th percentile points of each distribution, and show no significant correlation for the AB system, indicating a good orbital fit.
The orbital elements of the AB system are shown in Table~\ref{tab:orbs}.

With the updated orbital bundles generated for the AB system, we then took the measured AC astrometry as reported in Table 1 of \citet{Duchene03} and Table 1 of \citet{Boden12} and combined it with the 2021 astrometry.
We then corrected the AC relative astrometry using the measured masses of A and B and calculated the distance from C to the the barycentre of AB (called C-AB).
Using the C-AB astrometry, we carried out an {\tt orbitize!} fit with parameters identical to that of the AB system.
The orbital period of C around the AB system is on the order of hundreds of years, and so only a short arc of the orbit is traced out across the observed epochs.
The resultant orbital bundle shows a much wider spread of orbital parameters, ranging from periods of several decades to almost a thousand years.
The orbital bundle includes the combined mass of AB and C, including some unphysically low masses, so we keep bundles with masses between 3.5 and 7.0 $M_\odot$.

\begin{figure}[ht]
\begin{center}
    \centering
    \includegraphics[width=\columnwidth]{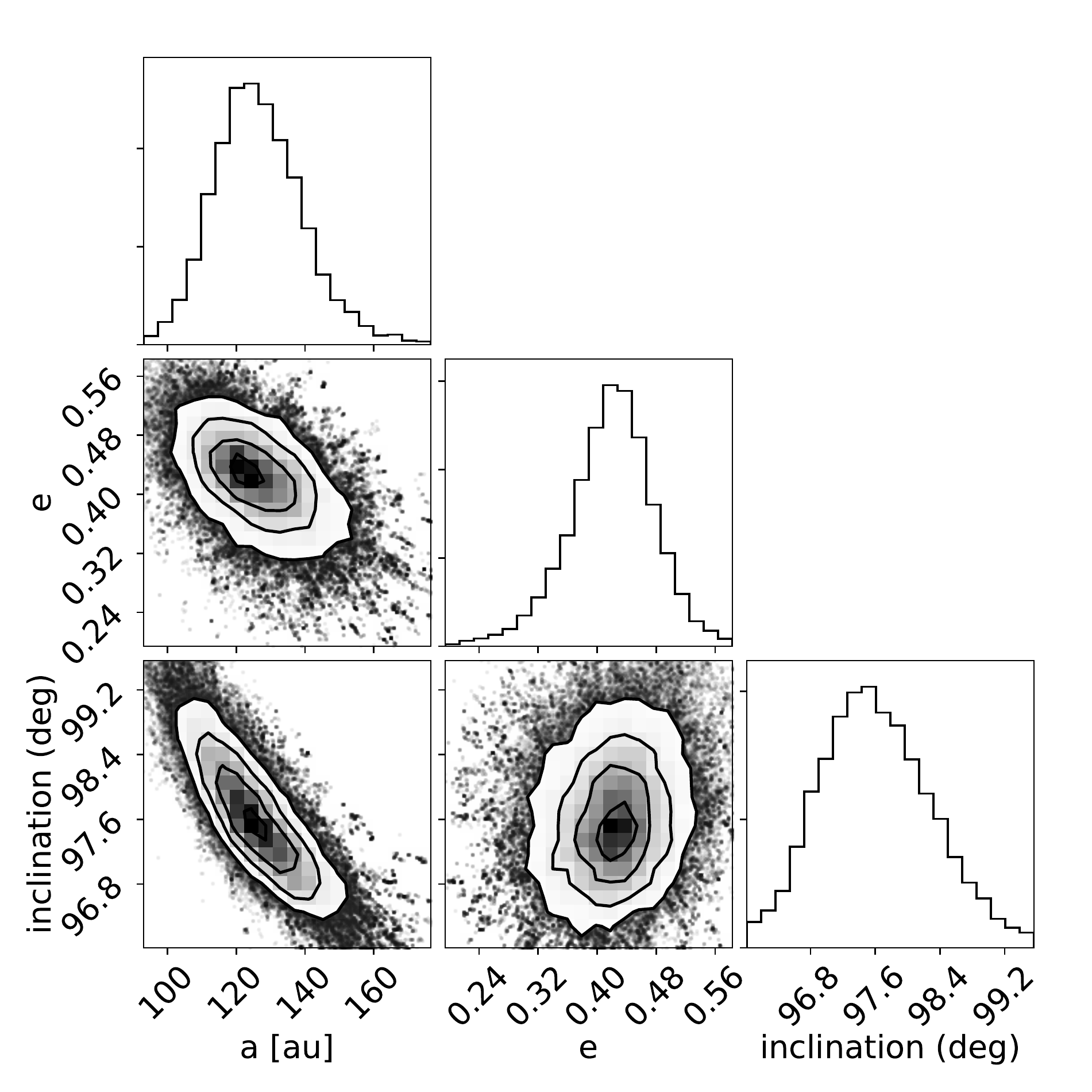}
        \caption{Correlation between the semi-major axis, eccentricity and orbital inclination of the V773 Tau C system.
        }
    \label{fig:v773caei}
    \script{analyze_and_plot_C_orbit_bundle.py}

\end{center}
\end{figure}

There is, however, the matter of orbital stability within a hierarchical triple, and we can approximate the V773 Tau system with A and B forming an inner binary and C the distant third component orbiting them.
Using the approximations within \citet{Eggleton95}, we can calculate a lower limit for the periastron of C around the AB system, and then reject orbits from the orbital bundle produced from our fitting procedure. This stability criteria is able to predict the stability over timescales of $5\times 10^5\,P_{AB}$ \citep{He2018}, much longer than the age of the system.
With $q_{in}=M_A/M_B=1.24$ and $q_{out}\geq 7.5$ and $a_{AB}=15.3au$ and $e_{AB} = 0.1$ this yields $Y^{min}_0=4.06$ and using Equation 2 from \citet{Eggleton95} we can derive a lower limit for the periastron distance of C.
Trial masses for C vary from 0.4 to 0.7 $M_\odot$ and only weakly change the periastron distance for C, so we set it at 68 au.
After removing orbits with periastron distances smaller than this, we use this new bundle of orbits as the starting point for a second optimization run.
The resultant orbital bundles show convergence and the triangle plot of the eccentricity and inclination of the orbit of C are very well constrained, as shown in Figure~\ref{fig:v773caei}, most notably that the inclination of C is $97\pm1$\degr{} with a tightly constrained eccentricity of 0.40.
The asymmetry of the errors on the orbital elements reflect the wide range of possible orbits as shown in Table~\ref{tab:orbs}.

\section{Rotational modulation and stellar activity} \label{sec:variability}

\begin{figure}
\begin{center}
    \centering
    \includegraphics[width=\columnwidth]{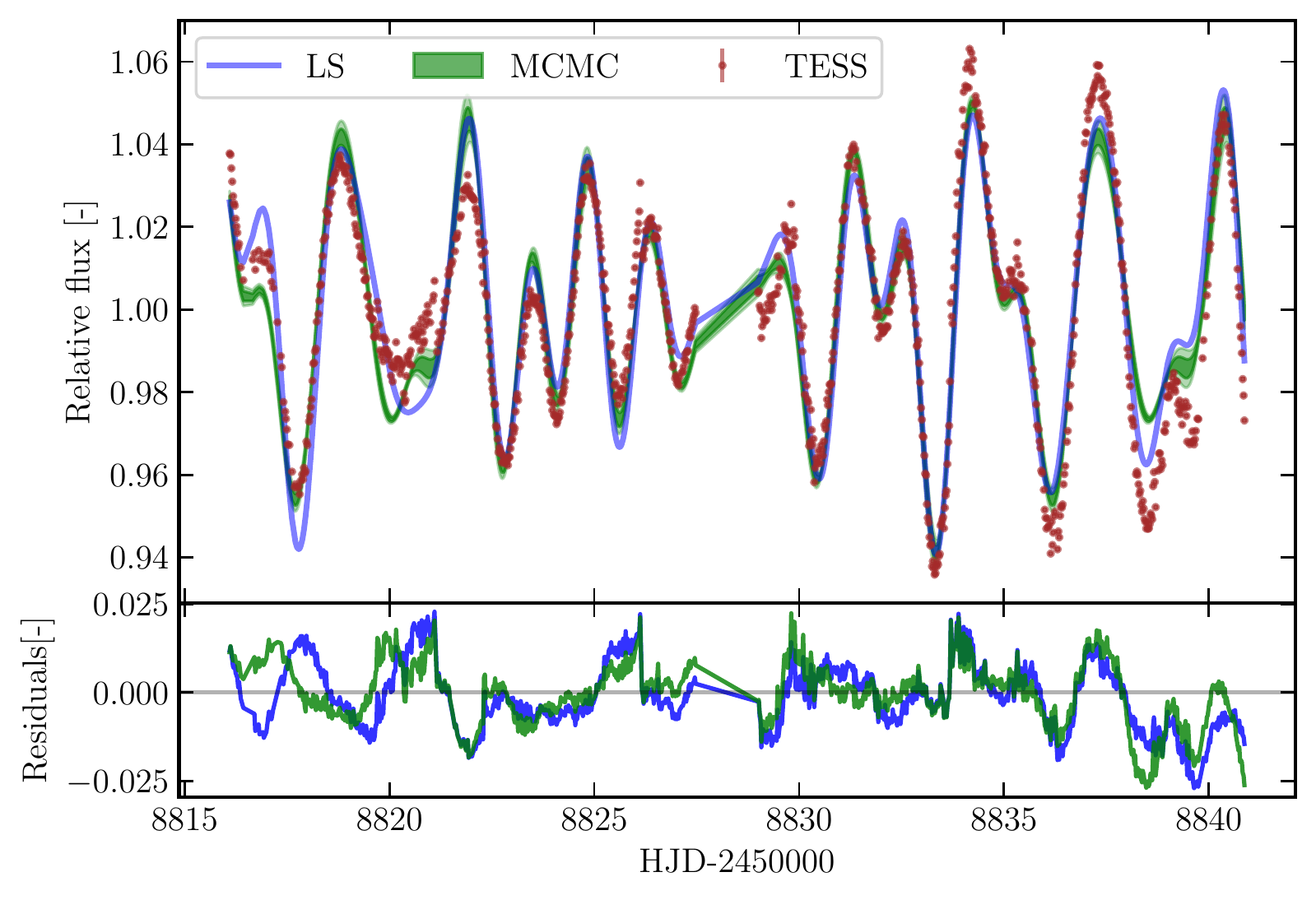}
    \caption{V773 Tau light curve as observed with TESS shown in the upper panel with brown dots.
    The Stellar Variability models computed with Lomb-Scargle (blue) and MCMC (green) are overlapped and the residuals are indicated as a color scale.}
\label{fig:tess_variability}
\end{center}
\end{figure}

\begin{figure}[ht]
    \centering
    \includegraphics[width=\columnwidth]{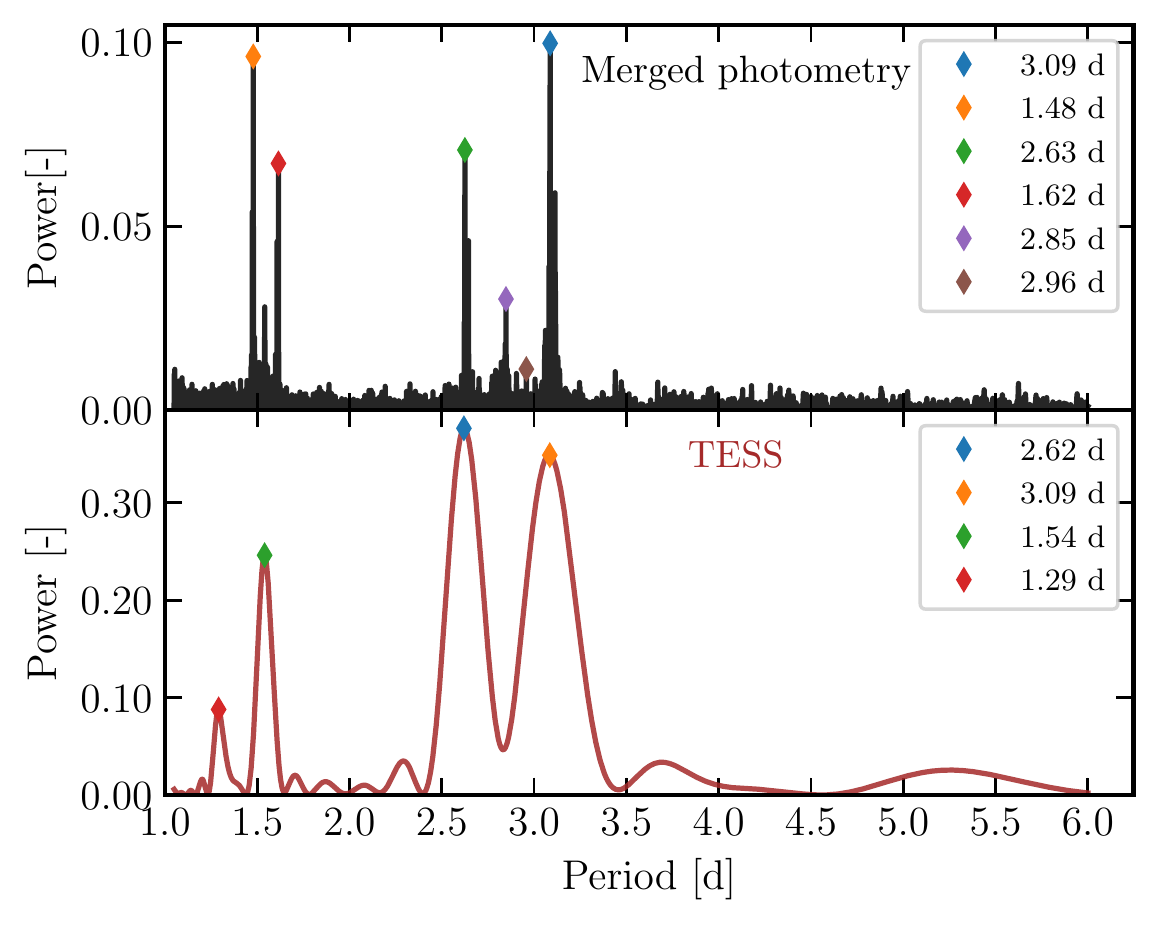}
    \caption{Lomb-Scargle Periodograms for the merged photometry (top) and TESS (bottom), the periods with the highest powers are highlighted.}
    \label{fig:two_periodograms}
    \script{fig9_periodograms.py}
\end{figure}

The V773 Tau system was observed by TESS in Sector 19 corresponding to MJD 58816 to 58842.
Several sinusoidal variations are seen in the light curve, with a total peak to valley variation of up to 10\%, as seen in Figure~\ref{fig:tess_variability}.
A Lomb-Scargle periodogram reveals four dominant frequencies in the power spectrum, with periods of 1.29, 2.62, 1.54 and 3.09 d.
Both stars are magnetically active, so it is reasonable to assume that there are multiple star spots present across their surfaces.
When combined with the rotation of the star, this leads to variations in observed flux as the spots rotate in and out of view.
We therefore interpret these periods as due to the rotational periods of 2.62 and 3.09 d for the stars Aa and Ab respectively. 
Since these variations are not pure sinusoids, this modulation appears at double the frequencies (half the periods) in the periodogram (see Figure~\ref{fig:two_periodograms}).

Rotation periods on the order of a few days are consistent with stellar evolution models at the age of the system, around 3 Myr.
Spectroscopic observations of chromospheric absorption lines show rotational broadening of $v \sin i = 38 \pm 4$ \kms{} for the Aa component \citep{Boden07}, and \citet{Welty95} quote velocities of $41.4$ \kms{} and $41.9$ \kms{} for Aa and Ab respectively.
Assuming stellar radii of $2.22\pm0.20$ $R_\odot$ and $1.74\pm0.19$  $R_\odot$ for Aa and Ab respectively \citep{Boden07}, we can calculate the expected rotational velocity for the two stars - with the two periods we see in the periodogram, we get $v_{Aa} = 36.4\pm 3.3$ \kms{} and $v_{Ab} = 34\pm 4$ \kms{}.
We derive the inclination of the rotational equator for both stars: $i_{Aa} = 73\pm 16\degr$ and $i_{Ab} = 53\pm 11\degr$ which, whilst not a strong constraint, are at least marginally consistent with the inclination of the AaAb orbital plane of $69\degr$.
Our assignment of the rotational periods to Aa and Ab could be incorrect - if these periods are assigned to the other star in the A system, then the star Aa has a rotational velocity equal to that of the projected rotational velocity, indicating that the star Aa is equator-on with an inclination of 90 degrees, and Ab is much closer to pole-on.
This would imply a far more dynamically misaligned system for A.

We model the light curve out of the eclipse as a sum of the four periods, mapped as a Fourier series of sines and cosines.
After removing these four periods, there remains a smaller amplitude sinusoidal signal with a period of 2.45 days.
We suggest that this is the rotational modulation for a star in the B system, but we regard this a somewhat tentative assignment that requires further confirmation with high spectral resolution spectroscopy to disentangle its nature.

\begin{figure}[ht]
    \centering
    \includegraphics[width=\columnwidth]{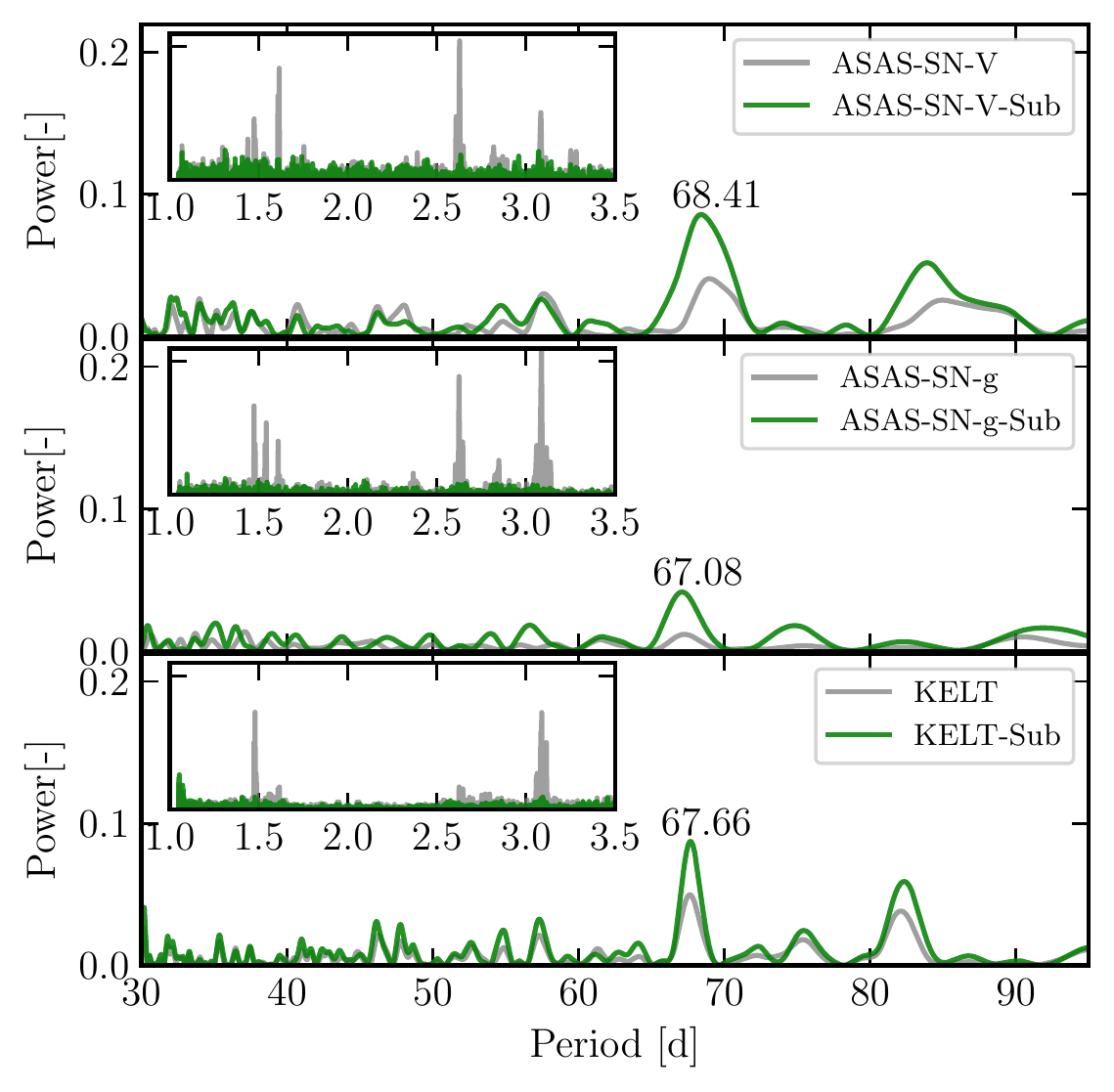}
    \caption{Lomb-Scargle Periodograms for the two bands of ASAS-SN and KELT.
    In grey (green) the periodogram before (after) removing the short-term rotational frequencies.
    A peak at $\sim 68$ days is highlighted for the three datasets.}
  \label{fig:long_periodogram}
  \script{fig10_long_periods.py}
\end{figure}

In addition to short-term modulation ($<10$ days), longer periodic signals can reveal the unknown orbital period of the B component and identify the reported $51$-day period of the A component.
Periodograms of the separate surveys with the short-term frequencies removed are shown in Figure~\ref{fig:long_periodogram}.
No 51 day period is seen, but there is a periodic signal at $\sim 68$ days which is clearly detected in all three photometric data sets.
This suggests that we may be seeing the orbital period for the B system, with the light curve modulation caused by the periodic change in orientation of the two stars.
An eccentric polar orbit in the B system would lift the apogee of both stars out of the midplane of the circumsecondary disk, and the resultant illumination change onto the upper surfaces of the disk leads to a periodic variation in the light curve of the system, and hence give rise to the observed signal in the periodogram.

\section{Disk model for the eclipse\label{sec:model}}

The V773 Tau system is unresolved by the ground-based survey telescopes, and so the light curves represent the summed flux of A, B and C.
The depth of the eclipse in the visible band is approximately 80\%.
The amount of flux from C at optical wavelengths is less than 1\% of the total flux, so we ignore the contribution of C in the photometric analysis.
A depth of 80\% is consistent at optical wavelengths with a complete eclipse of the Aa/Ab stellar components by an opaque occulter.
The compiled photometry covers a baseline of 18 years, and no other significant deep eclipse is seen, which rules out a dust cloud orbiting within the region of stability around the A system.

The orbit of B is inclined at 77\degr{} to our line of sight, and at the time of the eclipse, the B component was passing in front of the A component, close to the minimum projected separation between A and B (coincidentally just past the epoch of periastron).

We construct a toy model of the AB system with a disk around the B component in order to determine the possible parameters and geometry for a circumstellar disk that is consistent with the observed photometry.

\begin{figure}[ht]
\begin{center}
    \centering
    \includegraphics[width=\columnwidth]{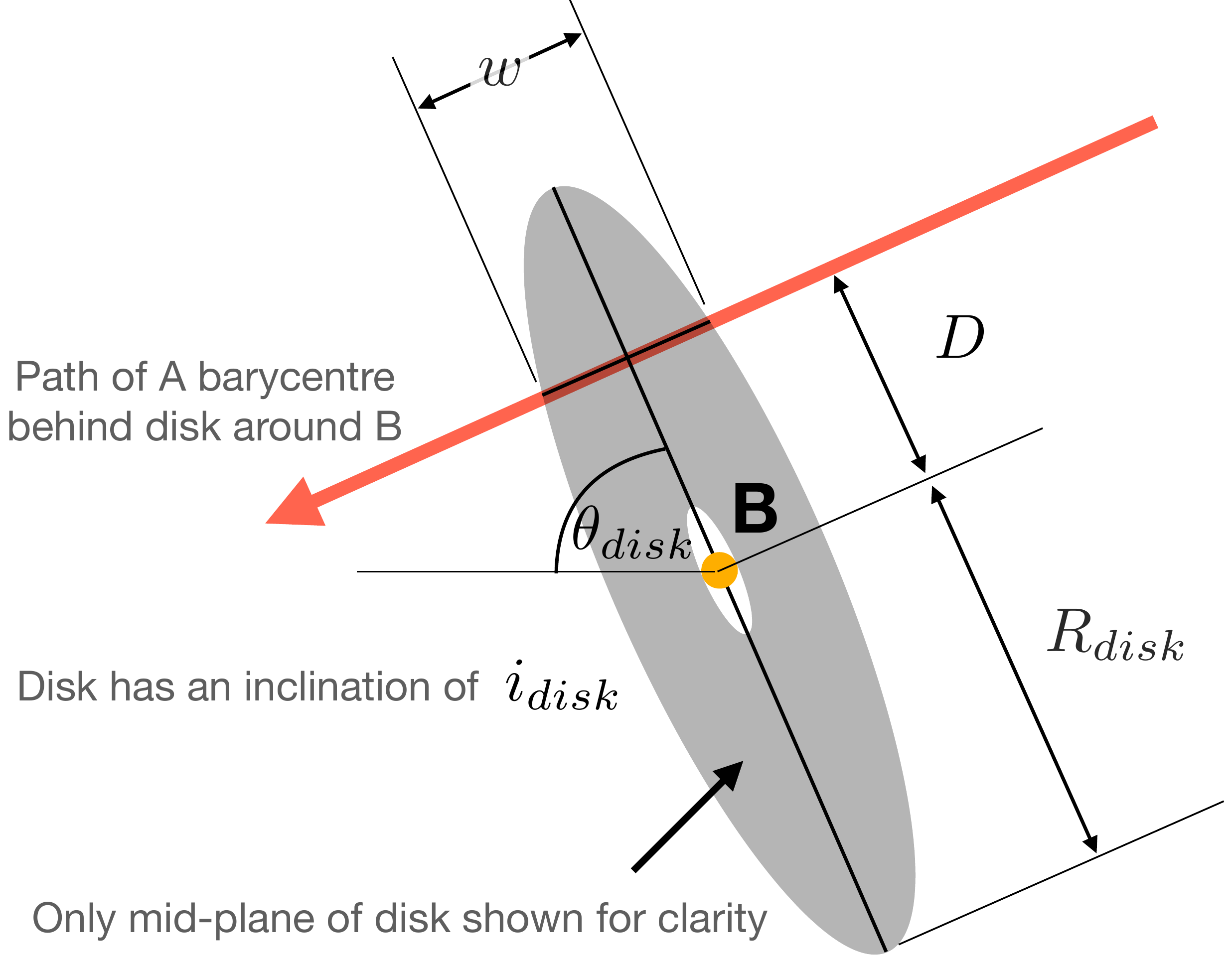}
    \caption{Sketch of the disk around B.
    The component A passes behind the disk with a projected chord length of $w$.
    Only the midplane of the disk is shown for clarity.
    The radius of the disk is $R_{disk}$ with an inclination to our line of sight of $i_{disk}$, and the disk is tilted by $\theta_{disk}$\degr{} anticlockwise from due East.}
\label{fig:csdcartoon}
\end{center}
\end{figure}

Circumstellar disks typically have a flared geometry, with a scale height for the dust that increases with increasing radii as a power law of the radius \citep{Dullemond02}.
The outermost parts of the disk will also be sculpted by the interaction with the A system and the competing torques from the B system and geometry of the AB orbit (discussed in Section~\ref{sec:theory}), resulting in many degrees of freedom for any fit to the photometry.
We simplify our model and approximate the disk as a cylindrical slab of dust with radius $R_{disk}$ with the density of dust being a function of the height above the midplane.
In cylindrical coordinates, we assume the absorbing material for $r<R_{disk}$ is $A(r,\phi,z)\propto z$ and $A=0$ for $r>R_{disk}$.
If the disk is exactly edge on to the line of sight then a single star passing behind the disk would have a light curve where the absorption is proportional to the dust absorption integrated along the line of sight from the background star to the Earth.
Since we are assuming a cylindrical slab, this integrated absorption is proportional to a Gaussian distribution, i.e. $A(z) = A_0 \exp(-(z/\sigma_{disk})^2)$ where $z$ is the vertical height above the midplane of the disk and $\sigma_{disk}$ is the characteristic scale height of the dust.
If the disk is tilted to the line of sight, then the star crosses a chord across the midplane of the disk of projected width $w$ (see Figure~\ref{fig:csdcartoon}), which samples a different part of the Gaussian distribution for that region of the disk.
The inclination of the disk with respect to the line of sight $i_{disk}$ is related to the radius and inclination of the disk by:

$$\sin i_{disk} = \frac{w}{2\sqrt{R_{disk}^2-D^2}} $$

The long axis of the projected disk is tilted at an angle $\theta_{disk}$\degr{} measured counterclockwise from due East, and the impact parameter for the path of A behind B is $D$.
Given the depth and duration of the eclipse, the extinction due to dust around B and the projected separation of A and B, we hypothesise that the disk is nearly edge on to our line of sight.

We determine the height of the stars Aa and Ab above the midplane of the edge-on disk around the B component by calculating the positions of the stars Aa, Ab and B as seen on the sky, calculating the relative positions of Aa and Ab from the B component, and then rotating these relative positions by an angle of $\theta_{disk}$ with the location of B as the origin into the coordinate frame of the disk.
We then calculate the height of the star above the disk as a function of time $t$, producing $z_{Aa}(t)$ and $z_{Ab}(t)$, and using the Gaussian function we calculate the flux of Aa and Ab through the disk a flux ratio of the two stars as $F(Ab)/F(Aa)$.

Our model disk flux is therefore:

$$F(t) = f( F(Ab)/F(Aa), w, \sigma_{disk}, A_{max}, \theta_{disk})$$ then represents the flux for both stars transmitted through a tilted disk with projected chord width $w$, the scale height $\sigma_{disk}$ and maximum absorption in the disk $A_{max}$.

The inclination of the disk is a function of both $w$ and the radius of the disk $R_{disk}$ which itself cannot be determined from the light curve.
Instead, we assume a given radius for the disk and then use $w$ to estimate the inclination of the disk.

We use the {\tt emcee} package to perform a fit of $F(t)$ to the photometric light curve over the epochs of the eclipse.
We minimise the $\chi^2$ statistic to obtain our fit, using 100 walkers and a burn in of 300 steps.
From \citet{Boden07}, the flux ratio between Aa and Ab was calculated to be $F(Ab)/F(Aa)=0.37\pm 0.03$ at a mean wavelength of 518 nm.
As a consequence, we carried out two separate model fits:
in the first case we fixed the flux ratio to be 0.37 and in the second case we let the flux ratio between the two A components be a free parameter.
The fits and marginalised errors on the parameters are shown in Table~\ref{tab:diskparams}.

\begin{table}
\caption{Parameters for the circumbinary disk model.}
\label{tab:diskparams}
\centering
\begin{tabular}{l c c}
\hline\hline 
Parameter                   & $F(Aa)/F(Ab)$ fixed          & All free \\
\hline 
Disk tip $\theta$ (deg)   & $52.87 \pm 0.04$    & $52.54 \pm 0.04$ \\
$A_{max}$                   & $0.878 \pm 0.03$    & $0.870 \pm 0.04$ \\
$F(Aa)$                     & $0.70$              & $0.35 \pm 0.02$ \\
$F(Ab)$                     & $0.30$              & $0.65 \pm 0.02$ \\
Chord width $w$ (mas)       & $6.78 \pm 0.08$     & $7.09 \pm 0.08$ \\
$\sigma_{disk}$ (mas)       & $1.59 \pm 0.04$     & $1.40 \pm 0.04$   \\
\hline
Chord width (au)         &   $0.900\pm0.011$ & $0.94\pm 0.01$             \\
$\sigma_{disk}$ (au)         &   $0.211\pm 0.005$ & $0.186\pm 0.005$             \\
Disk inclination (deg) &             $10.64\pm0.13$          &     $11.14\pm0.13$       \\
%
\hline                                             
\end{tabular}
\end{table}

%
%

A tilted disk around B can reach an outer radius of up to 0.38 $a_{AB}$, giving $R_{disk}=5.24$ au, for $e=0.1$ and an equal mass binary \citep[see Figure 4 from ][]{Miranda2015}.
The projected separation of A and B during the midpoint of the eclipse $D=4.78$ au, and the mean velocity of B around A as projected on the sky is $0.0111$ au d$^{-1}$.
We calculate that the circumbinary disk is close to edge on, approximately 10\degr{} from an edge on geometry.
An approximate measure of the flare of the disk is given by $\sigma_{disk}/D \approx 0.05$.
Finally, the inclination of the disk with respect to the AB orbital plane can take one of two values depending on whether the light curve ingress is due to the near side of the disk or far side of the disk, but both values are around 72\degr{} inclination, strongly supporting the theory that the disk around the B component is around a polar aligned binary with moderate eccentricity.


\begin{figure}[ht]
\begin{center}
    \centering
    \includegraphics[width=\columnwidth]{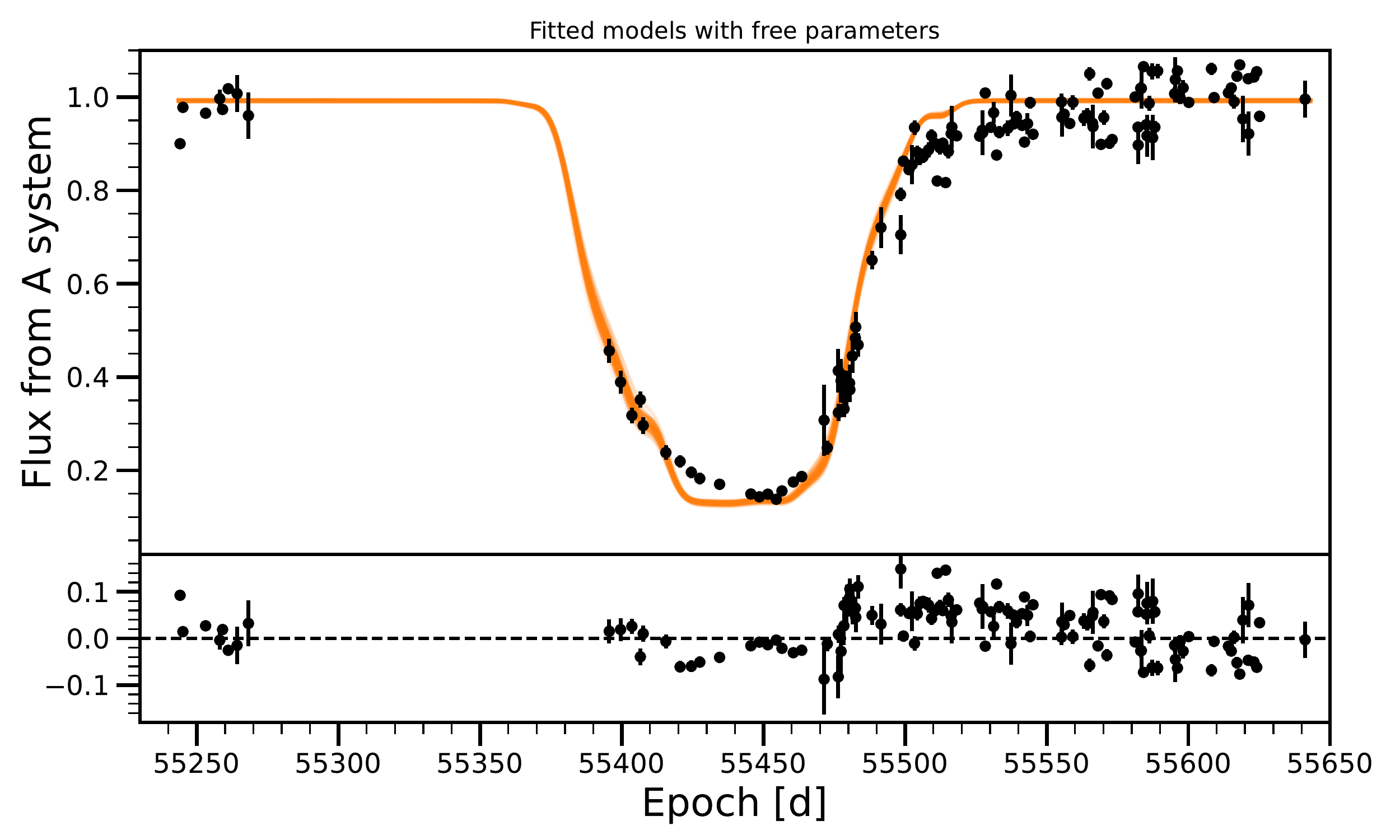}
    \caption{Model with all free parameters.}
\label{fig:fitallfree}
\end{center}
\end{figure}

\begin{figure}[ht]
\begin{center}
    \centering
    \includegraphics[width=\columnwidth]{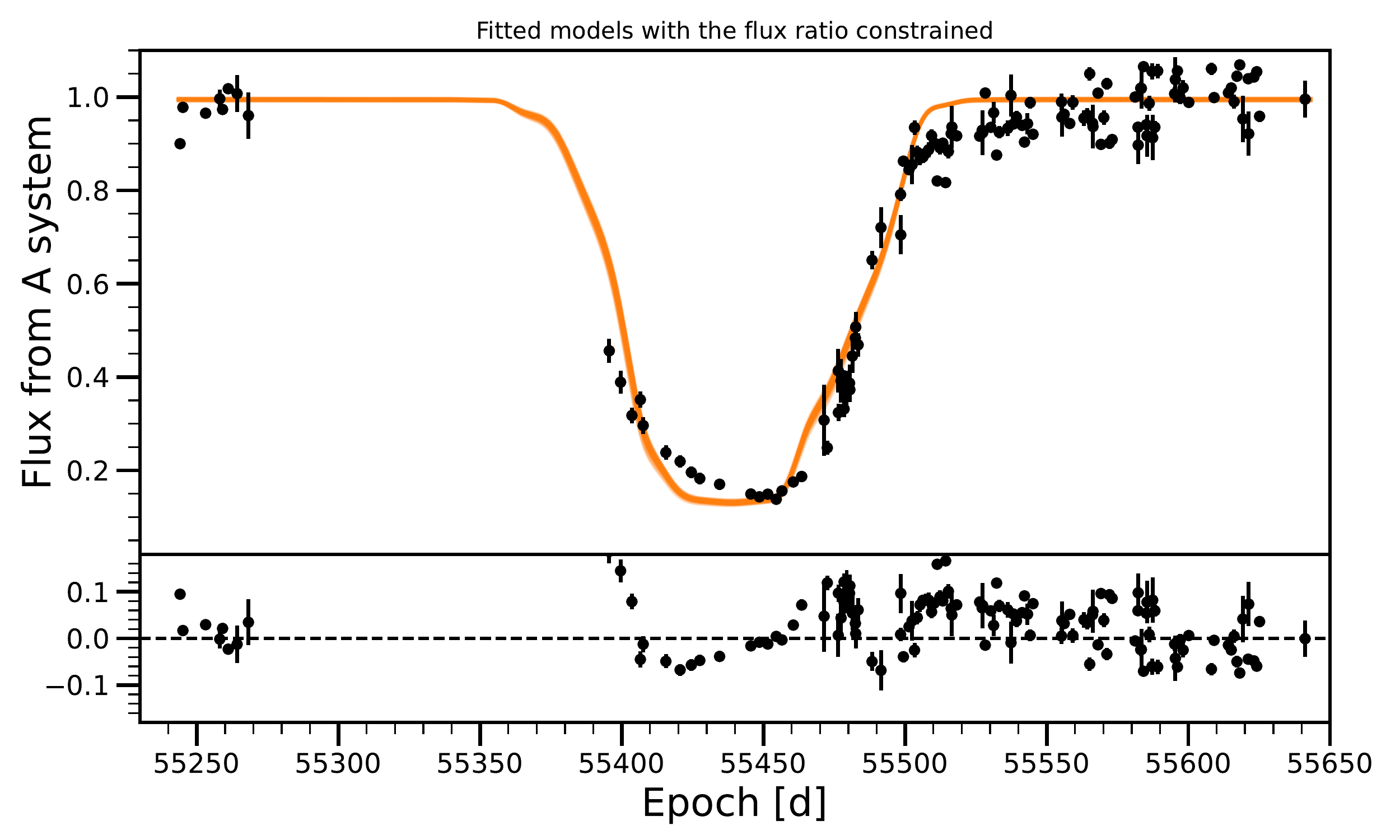}
    \caption{Model with $F(Aa)/F(Ab)$ constrained.}
\label{fig:fitconstrained}
\end{center}
\end{figure}

\section{A tilted polar circumbinary disk}\label{sec:theory}

In this Section we consider the dynamics of a circumbinary disk in this quadruple star system. 
In order for the stellar system to be stable, each binary must be inclined by less than about $40^\circ$ to the AB binary orbital plane otherwise, the stars will undergo Kozai-Lidov (KL) oscillations of eccentricity and inclination \citep{vonZeipel1910,Kozai1962,Lidov1962}.
A stable circumbinary disk around stars undergoing KL oscillations is unlikely. 
The disk, however, is likely to  be above the critical KL inclination since is it close to $90^\circ$ to the AB binary orbital plane \citep{Lubow2017,Zanazzi2017}.
If the disk were orbiting a single star, it would be KL unstable \citep{Martin2014,Fu2015} and would move towards alignment with the AB binary orbital plane on a timescale of tens of orbital periods of the AB binary.
However, the inner binary causes nodal precession that can stabilise the disk against KL oscillations \citep{Verrier2009,Martin2022} and lead to polar alignment in which the disk is perpendicular to the binary orbit and aligned to the eccentricity vector of the inner binary \citep{Martin17}.

\subsection{Particle dynamics}

\begin{figure}
\begin{center}
    \centering
    \includegraphics[width=\columnwidth]{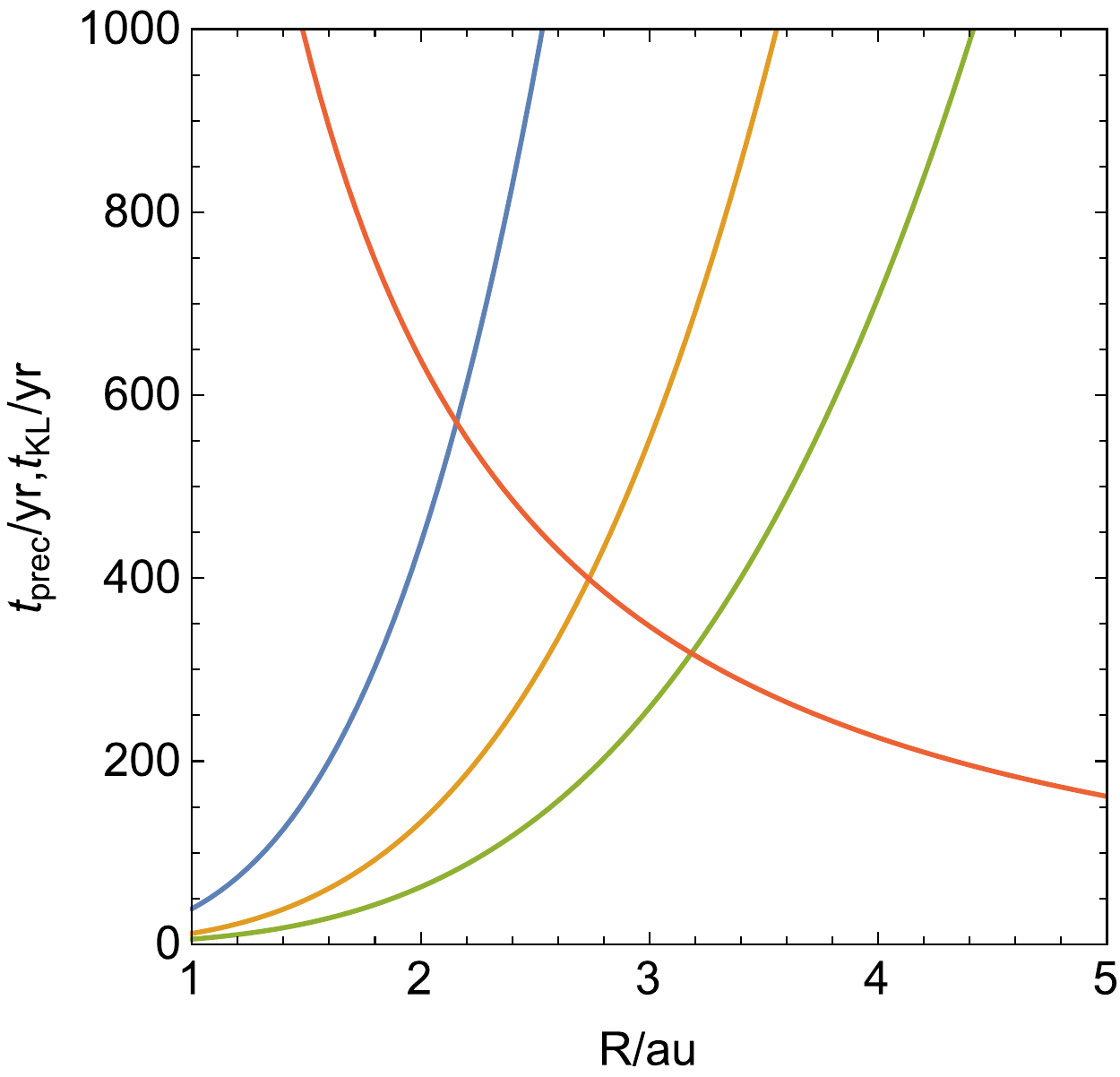}
    \caption{Particle nodal precession timescale and the KL timescale as a function of the particle separation from  the BaBb binary. 
    The blue ($e_{\rm b}=0.2$),  orange ($e_{\rm b}=0.5$) and green ($e_{\rm b}=0.8$) lines show the nodal precession timescales.
    The red line shows the KL timescale.
    The particle is unstable if  the nodal precession timescale is longer than the KL timescale \citep{Verrier2009}. }
\label{fig:particles}
\end{center}
\end{figure}

We consider the dynamics of a test particle that orbits the BaBb binary.
First, we examine the effect of the inner binary on the test particle in the absence of the outer binary.
The particle is highly inclined to the inner binary orbit.
It undergoes nodal precession about the eccentricity vector of the binary \citep{Verrier2009,Farago2010,Doolin2011,Chen2019}.
The frequency for the precession of a particle at semi-major axis $R$ is given by
\begin{equation}
    \omega_{\rm p}=\frac{3}{4}k \frac{M_{Ba}M_{Bb}}{M_{BaBb}^2}
 \left(\frac{a_{BaBb}}{R}\right)^{7/2} \Omega_{BaBb},
\end{equation}
where the B binary angular frequency given by
\begin{equation}
\Omega_{BaBb}=\sqrt{\frac{G M_{BaBb}}{a_{BaBb}^3}}.
\end{equation}
and
\begin{equation}
    k=\sqrt{5} e_{BaBb} \sqrt{1+4e_{BaBb}^2}
\end{equation}
\citep{Farago2010,Lubow2018}. The timescale for the precession is
\begin{equation}
    t_{\rm prec}= \frac{2\pi}{\omega_{\rm p}}.
\end{equation}
We assume that the inner binary has equal mass components with a total mass of $M_{BaBb}=2.4\,\rm M_\odot$.
The binary orbits with a semi-major axis of $a_{BaBb}=0.43\,\rm au$.
Figure~\ref{fig:particles} shows the nodal precession timescale for particles around binaries with three different eccentricities.
The nodal precession timescale increases with distance from the inner binary and decreases with binary eccentricity. 

In the absence of the inner binary, the outer binary component causes KL oscillations of the highly inclined test particle.
These are oscillations in the eccentricity and inclination of the orbit \citep[e.g.][]{Naoz2016}.
These occur on a timescale given by
\begin{equation}
    t_{\rm KL}\approx \frac{M_{AaAb}+M_{BaBb}}{M_{AaAb}}\frac{P_{AB}^2}{P_{\rm p}}
\end{equation}
for a circular orbit outer binary \citep{Kiseleva1998,Ford2000}, where the orbital period of the particle is $P_{\rm p}=2\pi/\sqrt{GM_{BaBb}/R^3}$.
The outer binary companion has mass  $M_{AaAb}=2.9\,\rm M_\odot$ and is in a circular orbit with semi-major axis $a_{AB}=15\,\rm au$.
The red line in Figure~\ref{fig:particles} shows the KL timescale.

Particles are unstable outside of the radius where the KL timescale becomes smaller than the nodal precession timescale \citep{Verrier2009,Martin2022}.
Thus, in the absence of gas, solid particles are only stable to a  radius much smaller than $5\,\rm au$.
The larger the inner binary eccentricity, the farther out stable particles can exist.
However, since the disk in V773 Tau is observed to extend to a radius of around $5\,\rm au$, there must be gas present in the disk.
A gas disk is in radial communication and this can allow it to be stable in a region that individual particles may be unstable.

\subsection{Disk dynamics}

The outer edge of the disk is likely tidally truncated by the torque from the A binary component.
The outer radius of the disk decreases with the eccentricity of the AB binary \citep{Artymowicz1994} and the inclination of the disk to the AB binary orbital plane \citep{Lubow2015}.
The disk can extend to a radius of about $0.38\,a_{AB}$ when it is misaligned by $90^\circ$ to the AB binary orbital plane \citep{Miranda2015}.

If the disk is in good radial communication, it can undergo solid body precession  \citep{Larwood1996}.
Radial communication in the disk is maintained through pressure induced bending waves that travel at a speed of $c_{\rm s}/2$ \citep{PapaloizouLin1995,Lubow2002}, where $c_{\rm s}\approx H\Omega$ is the gas sound speed.
The radial communication timescale is $t_{\rm c}\approx 2 R_{\rm out}/c_{\rm s}$.
With $H/R=0.05$ and $R_{\rm out}=5\,\rm au$ we have $t_{\rm c}=47\,\rm yr$.
Provided that the precession is on a timescale that is significantly longer than this, then we expect solid body precession of the disk.
Note that if the disk is not in good radial communication then it may undergo breaking where the disk forms disjoint rings that precess at different rates \citep[e.g.][]{Larwood1996,Nixon2013}.
This may shorten the alignment timescale \citep[e.g.][]{Smallwood2020}.

\begin{figure}
\begin{center}
    \centering
    \includegraphics[width=\columnwidth]{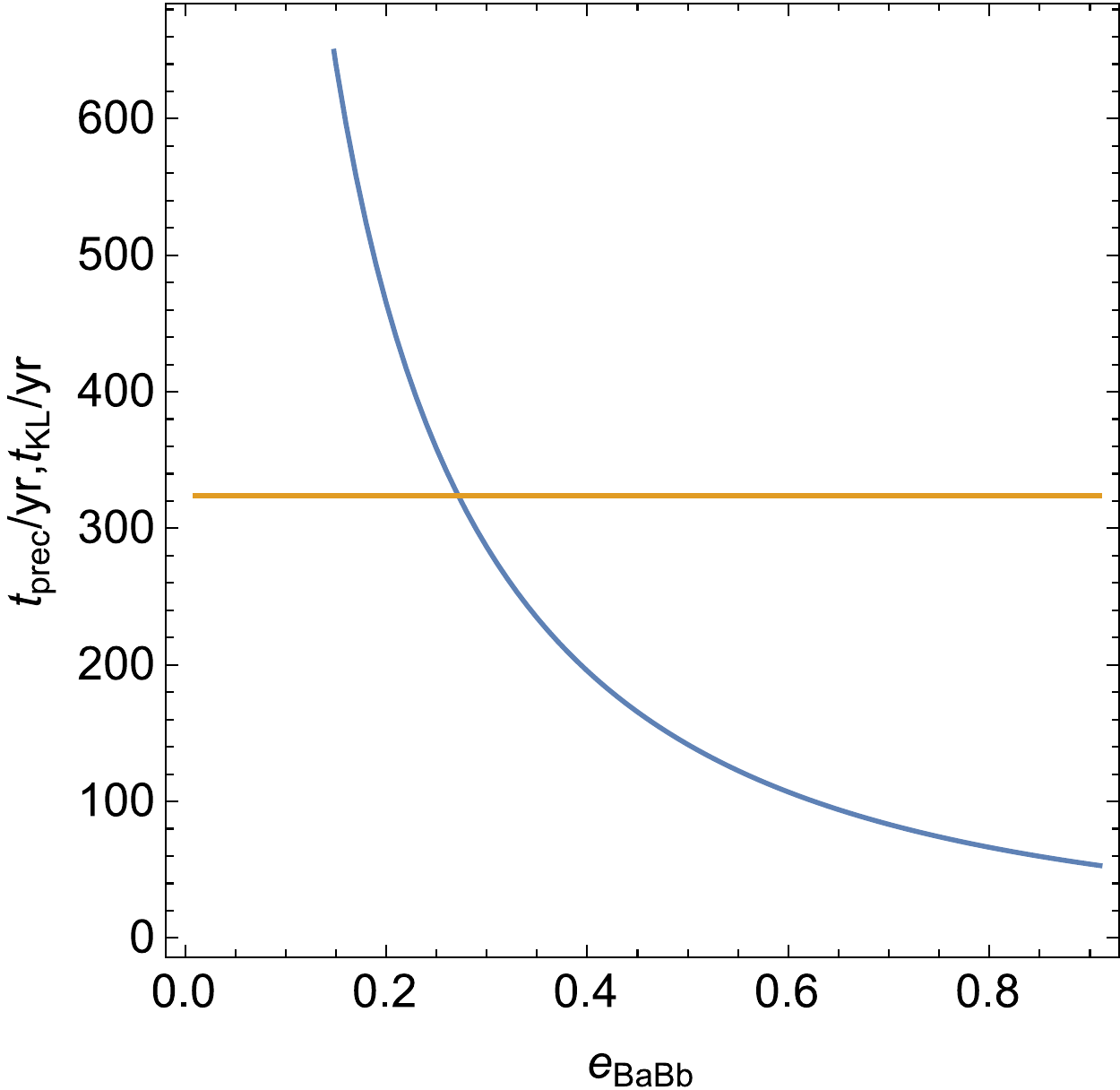}
    \caption{The disk nodal precession timescale (blue) and the KL timescale (orange) as a function of the eccentricity of the BaBb binary.
    The disk can remain polar if the nodal precession timescale is shorter than the KL timescale.  }
\label{fig:timescales}
\end{center}
\end{figure} 

We first ignore the A binary and consider the dynamics as a result of the inner binary.
For a sufficiently high inclination, the disk angular momentum vector precesses about the binary eccentricity vector \citep{Aly2015} and because of viscous dissipation aligns towards this polar inclination \citep{Martin17}.
We consider the precession rate for the circumbinary disk as a result of the torque from the inner binary for varying inner binary separation.
 The inner binary carves a cavity in the inner parts of the disk.
 The size of the cavity depends upon the binary eccentricity and the disk inclination \citep{Artymowicz1994,Miranda2015,Franchini19}.
 We assume that the surface density extends from an inner radius of $R_{\rm in}=2.5\,a_{BaBb}$ out to $R_{\rm out}=5\,\rm au$ with a profile  $\Sigma \propto R^{-3/2}$. 
 
We calculate a disk density averaged nodal precession timescale with equation~(16) in \cite{Lubow2018}.
The blue line in Figure~\ref{fig:timescales} shows the circumbinary disk nodal precession timescale for a disk that is precessing about the BaBb binary eccentricity vector.
We also calculate the disk density weighted KL timescale with equation~(4) in \cite{Martin2014} and this is shown in the yellow line.
The disk KL oscillation timescale is shorter than the nodal precession timescale only for small binary eccentricities.
We expect that a disk may remain in a stable polar configuration for binary eccentricity $e_{\rm b}\gtrsim 0.3$.
However, we caution that this should be verified with numerical simulations that include both binary components.
While a particle that undergoes KL oscillations becomes unstable because of the high eccentricity achieved, the KL oscillations of a disk are damped and do not reach such high eccentricities \citep[e.g.][]{Fu2015,Fu2015b}.
Further, disk breaking may be more likely with both an inner and an outer binary and so an inner polar ring may still be possible even around smaller binary eccentricity \citep{Martin2022}.
However, in order to have the disk out to $5\,\rm au$ in a polar configuration, we suggest that a larger binary eccentricity is required.

Note that all of the timescales calculated in this section are approximate since we assume a truncated power law surface density.
The surface density profile is affected by the inclination of the disk and will evolve as the disk aligns.
Hydrodynamical simulations of a polar aligning disk found that the alignment timescale was about a factor of two faster in the simulation as predicted by the linear theory \citep{Smallwood2020}.
This system should be investigated in hydrodynamical simulations in the future.
This would allow us to put stronger constraints on the orbital parameters of the BaBb binary.

\section{Discussion}\label{sec:discuss}

\subsection{Orbital dynamics of the C component}


Without radial velocity measurements, orbital fits with direct imaging have two degenerate orbital solutions.
For the orbit of C around the AB system, one orbital solution has the orbital vector almost antiparallel to the orbit of the AB system (where C is in front of the AB system) at 166\degr{}.
The other solution has C behind the AB system and with C orbiting in the same direction as the AB system, with a mutual inclination of 29\degr{}.
If the stellar system formed from the same cloud of protostellar material, it suggests that the inclination of 29\degr{} is the correct orbit, but radial velocity measurements of C will break this degeneracy.

\subsection{The B circumbinary disk}

The eclipse cannot be fit with an exactly edge on Gaussian disk - the bottom of the eclipse is too broad with respect to the wings of the eclipse.
A disk that is tilted with respect to the line of sight provides a significantly improved fit, but without a measured outer radius for the disk, there is a degeneracy between the tilt of the disk and the radius.
We ran two separate models for the eclipse - in the first one we fix the brightness ratio of the A binary at the measured value and for the second model we leave it as a free parameter - we note that the free parameter fit gives a brightness ratio for Aa and Ab that is almost exactly the opposite of that measured in previous papers.
The free parameter fit had a lower $\chi^2$ value, but both models yielded similar parameters for the disk inclination, chord width and scale height of the disk.
Possible explanations for this include an incorrect assumption about the symmetry of the disk, especially since we are probing radii close to the outer edge of the disk where warping from coplanar geometry is possible.
Another degeneracy exists with the direction of the tilt - whether the leading edge of the eclipse is caused by the front edge or the back edge of the disk.
In either possible configuration, the disk has a significant inclination of around 70\degr{} with respect to the orbital plane of AB.

\subsection{The next eclipse}
The resulting orbital fit bundle is shown in Figure~\ref{fig:v773orbitize}.
We can see that using the astrometric fit that the midpoint of the next eclipse will be at 2037 March 10 plus or minus 25 days ($2037.19\pm0.07$\ yr).
As the next eclipse approaches, nightly photometric monitoring with the AAVSO association will alert observers to start a detailed observational campaign.

\section{Conclusions}\label{sec:conclusions}

We have discovered an extended eclipse seen towards the V773 Tau multiple star system, and we hypothesise that it is due to a disk of dust orbiting around the B component.
In order to see the eclipse, the disk must be significantly inclined with respect to the orbital plane of the AB system, implying a restoring torque that prevents the disk from becoming coplanar with the AB orbit.
The V773 Tau system has a large mid-IR flux \citep{Prusti92,Duchene03,2007AAS...211.2904P}  above that of the photospheric levels expected from the stellar components, and sub-mm flux at $850 \mu m$ and 1.3 mm \citep{Andrews05} consistent with the presence of dust in the system.
Due to the confounding effects of the dust around C \citep{Duchene03,Woitas03}, estimating a mass of the disk around B is not possible without spatially resolved sub-mm imaging to distinguish separate dust emission contributions from each source.

Several observations indicate that the B component is itself a binary system: the luminosity observed for B is consistent with that from two equal mass components, the inclined disk can be dynamically stable if the B binary has a moderately eccentric orbit, and a 67 day modulation in the light curve of the system would be consistent with a change in illumination of the circumbinary disk due to the eccentric orbital motion of the two B components in a polar orbit.
We conclude that B is a moderately eccentric, nearly equal mass binary on a 67 day orbital period with polar orientation with respect to the surrounding disk.

A direct imaging observation of the V773 Tau system enables a refinement of the orbital elements of the AB system and a first determination of the orbital elements of the AB-C system.
The C system has a well constrained inclination and eccentricity, with an orbital period from 570 to 700 years, and is moderately inclined $\approx 30$\degr{}) with respect to the AB system.

The fortuitous alignment of the circumsecondary disk around V773 Tau B enables additional characterisation of a young multiple stellar system that contains three components at different stages of evolution.
An observational campaign during the next eclipse in 2037 will enable spectroscopic characterisation of the disk, including gas and dust kinematics at scales significantly smaller than that can be observed with large distributed interferometers such as the Atacama Large Millimeter/submillimeter Array.

\begin{acknowledgements}

We thank our referees for their constructive comments which helped improve this paper.
This research has used the SIMBAD database, operated at CDS, Strasbourg, France \citep{wenger2000}.
This work has used data from the European Space Agency (ESA) mission {\it Gaia} (\url{https://www.cosmos.esa.int/gaia}), processed by the {\it Gaia} Data Processing and Analysis Consortium (DPAC, \url{https://www.cosmos.esa.int/web/gaia/dpac/consortium}).
Funding for the DPAC has been provided by national institutions, in particular the institutions participating in the {\it Gaia} Multilateral Agreement.
To achieve the scientific results presented in this article we made use of the \emph{Python} programming language\footnote{Python Software Foundation, \url{https://www.python.org/}}, especially the \emph{SciPy} \citep{virtanen2020}, \emph{NumPy} \citep{numpy}, \emph{Matplotlib} \citep{Matplotlib}, \emph{emcee} \citep{foreman-mackey2013}, and \emph{astropy} \citep{astropy_1,astropy_2} packages.
We acknowledge with thanks the variable star observations from the AAVSO International Database contributed by observers worldwide and used in this research.
We thank the Las Cumbres Observatory and its staff for its continuing support of the ASAS-SN project, and the Ohio State University College of Arts and Sciences Technology Services for helping us set up and maintain the ASAS-SN variable stars and photometry databases.
ASAS-SN is supported by the Gordon and Betty Moore Foundation through grant GBMF5490 to the Ohio State University and NSF grant AST-1515927.
Development of ASAS-SN has been supported by NSF grant AST-0908816, the Mt. Cuba Astronomical Foundation, the Center for Cosmology and AstroParticle Physics at the Ohio State University, the Chinese Academy of Sciences South America Center for Astronomy (CASSACA), the Villum Foundation, and George Skestos.
Early work on KELT-North was supported by NASA Grant NNG04GO70G.
Work on KELT-North was partially supported by NSF CAREER Grant AST-1056524 to S. Gaudi.
Work on KELT-North received support from the Vanderbilt Office of the Provost through the Vanderbilt Initiative in Data-intensive Astrophysics.
Part of this research was carried out in part at the Jet Propulsion Laboratory, California Institute of Technology, under a contract with the National Aeronautics and Space Administration (80NM0018D0004).
This publication makes use of VOSA, developed under the Spanish Virtual Observatory project supported by the Spanish MINECO through grant AyA2017-84089.
VOSA has been partially updated by using funding from the European Union's Horizon 2020 Research and Innovation Programme, under Grant Agreement number 776403 (EXOPLANETS-A)
The research of CA is supported by the Comit\'e Mixto ESO-Chile and the VRIIP/DGI at the University of Antofagasta.
G.M.K. is supported by the Royal Society as a Royal Society University Research Fellow.
M.A.K. thanks Sarah Blunt and Jason Wang for discussions on setting up orbit bundles.

\end{acknowledgements}

\bibliography{bib}

\end{document}